\documentclass[apj,twocolumn, floatfix]{aastex701}
\usepackage{amsmath}
\usepackage{longtable}
\usepackage{graphicx}
\usepackage{ulem}
\usepackage{xcolor}
\usepackage{lettrine}
\usepackage{yfonts}
\usepackage{rotating} 
\definecolor{meridithgreen}{RGB}{0, 150, 0}

\newcommand{\GAs}{GAs}

\usepackage[colorinlistoftodos]{todonotes}

\shorttitle{Bulge GCE}
\shortauthors{Miller et al.}

\begin{document}







\title{The Two-infall Model Revisited: Constraints on Milky Way Bulge Assembly\\from $>$30,000 Galactic Chemical Evolution Models and Machine Learning}

\author[0000-0002-3780-0592]{Niall Miller}
\affiliation{University of Wyoming, 1000 E University Ave, Laramie, WY 82071, USA}
\email{niall.j.miller@gmail.com}

\author[0000-0002-8717-127X]{Meridith Joyce}
\affiliation{University of Wyoming, 1000 E University Ave, Laramie, WY 82071, USA}
\email{mjoyce8@uwyo.edu}

\author[0000-0002-8878-3315]{Christian I. Johnson}
\affiliation{Space Telescope Science Institute, 3700 San Martin Drive, Baltimore, MD 21218, USA}
\email{chjohnson1@stsci.edu}

\author[0000-0002-4818-7885]{Jamie Tayar}
\affiliation{Department of Astronomy, University of Florida, Gainesville, FL 32611, USA}
\email{jntayar@gmail.com}

\author{Thomas Trueman}
\affiliation{University of Bayreuth, BGI, Universitätsstraße 30, 95447 Bayreuth, Germany}
\affiliation{Konkoly Observatory, HUN-REN Research Centre for Astronomy and Earth Sciences, Konkoly Thege Miklós út 15-17, H-1121 Budapest, Hungary}
\affiliation{CSFK, MTA Centre of Excellence, Budapest, Konkoly Thege Miklós út 15-17, H-1121 Budapest, Hungary}
\email{tomtrueman@hotmail.com}

\author[0000-0003-0427-8387]{R. Michael Rich}
\affiliation{Department of Physics and Astronomy, UCLA, PAB 430 Portola Plaza, Box 951547, Los Angeles, CA 90095-1547, USA}
\email{rmrastro@gmail.com}


\begin{abstract}
We constrain the formation history of the Milky Way bulge using a two-infall galactic chemical evolution (GCE) algorithm implemented in the N'OMEGA+ code.
We recover a best-fit scenario in which the bulge forms through an early, rapid starburst ($t_1 \sim 0.1$~Gyr, $\tau_1 \sim 0.09$~Gyr, and star formation efficiency (SFE) $\sim 3~{\rm Gyr}^{-1}$), followed by a delayed, lower-mass second infall ($t_2 \sim 5.1$~Gyr, $\tau_2 \sim 1.7$~Gyr, and $\sigma_2 \sim 0.69$).
Our model adopts mass- and metallicity-dependent nucleosynthetic yields from modern stellar grids and explores a wide GCE parameter space in infall timing, SFE, mass partitioning, initial mass function upper mass, and type Ia supernova normalization, optimized via a hybrid genetic algorithm with Markov Chain Monte Carlo refinement.
The later infall features a reduced SFE ($\Delta{\rm SFE} \sim 0.72$), reproducing the metal-rich peak of the bulge metallicity distribution function (MDF) and the decline in [$\alpha$/Fe] at high [Fe/H]. Our model naturally favors the M. Joyce et al. age--metallicity relation over the ages in T. Bensby et al.
Degeneracy and principal component analyses show that the infall history, SFE, and mass partitioning are strongly covariant---the bulge's observed MDF, abundance trends, and age distribution constrain only their combinations, not each parameter independently.
The results support a composite bulge origin---an early, rapid collapse builds the majority of the mass, while a younger component is required to match the late-stage enrichment.
\end{abstract}


\section{Introduction}

The Galactic bulge remains an archetype of composite stellar populations and overlapping formation channels.
The bulge's formation history is complex and multifaceted, with various scenarios proposed in the literature.
Previous works have suggested that the bulge formed primarily through a classical mechanism involving early mergers of primordial structures in a Lambda cold dark matter ($\Lambda$CDM) context \citep{Ortolani95, Baugh96, Abadi03a, Abadi03b}.
Others have argued for a secular origin, where the bulge arises from disk instabilities and bar formation \citep{Combes90, Raha91, ONeill03, Athanassoula05, Shen10, Ciambur21, Ghosh23}.
More recent studies have proposed hybrid scenarios, combining an initial rapid collapse with later bar-driven evolution \citep{McWilliam10, Wegg13, Ness16, Barbuy18}.
Barred regions often exhibit flattened age and metallicity gradients compared to their disks \citep{Seidel2016, Fraser2020, Neumann2020}.

Structurally, the bulge shows evidence of both a boxy/peanut bar and a classical spheroid \citep{Weiland94, McWilliam10, Wegg13, Ness16, Barbuy18}. However, more recent dynamical studies find that any classical spheroid component contributes at most a few percent of the total stellar mass of the central Milky Way \citep{Wegg2015, Portail2017}, with even the high-$\alpha$ population exhibiting cylindrical rotation consistent with the bar.
Chemically, the bulge exhibits a broad metallicity distribution with distinct $\alpha$-enhancement patterns, motivating two-component formation scenarios \citep{Babusiaux10, Bensby11, Gonzalez11, Hill11, Uttenthaler12}.
The metallicity distribution function (MDF) of the bulge is characterized by a multipeak shape, often bimodal, with a metal-poor peak centered at [Fe/H] $\sim -0.3$ dex and a metal-rich peak at [Fe/H] $\sim +0.3$ dex \citep{Bensby11, Hill11}.
This bimodality is also reflected in the [$\alpha$/Fe] versus [Fe/H] abundance ratios, suggesting the presence of different stellar populations---a metal-poor population with historically attributed spheroid kinematics and enhanced [$\alpha$/Fe] ratios, indicative of rapid formation, and a metal-rich one with bar-like kinematics and near-solar [$\alpha$/Fe], possibly originating from the inner disk \citep{Matteucci90, Rich90, Rich1990, McWilliam94, Minniti1996, Valenti05}. We note, however, that recent kinematic analyses suggest even the high-$\alpha$ population may share cylindrically rotating, bar-like kinematics rather than representing a distinct spheroidal component \citep{Wegg2015}. 

The bulge's MDF bimodality, the split in its $\alpha$-element trends, and the coexistence of old $\alpha$-enhanced stars with a younger, metal-rich population all imply at least two chemically distinct enrichment phases. 
These observational features cannot be reproduced by a single rapid star formation event, motivating models with multiple, temporally separated gas accretion episodes \citep[e.g.,][]{Matteucci19, Molero2024}.
Modeling the bulge formation history as a two-infall process is a natural mechanism for exploring these complexities. 
While the two-infall paradigm was originally developed in the context of the Galactic disk, its conceptual structure remains useful when applied to the bulge, and so we draw on those disk-based studies for methodological guidance.
In this paradigm, an early rapid infall (producing $\alpha$-rich stars) is followed by a slower, prolonged phase of star formation (producing more Fe-rich stars).
Subsequent studies refined the two-infall scenario to reproduce abundance data across the disk (e.g., \citealt{Chiappini2001, Chiappini2003, Griffith2021}).
Others have explored the supernova Ia (SN\,Ia) delay-time distribution \citep[DTD;][]{Matteucci09, Palicio2023B}, galactic fountains \citep{Spitoni09}, radial gas flow variations in the star formation efficiency \citep[SFE;][]{SpitoniMatteucci11}, radial stellar migration \citep{Morossi15}, azimuthal abundance variations \citep{Spitoni2019}.
\citet{Palla24} proposed a third gas accretion event in the last $\sim 3$ Gyr to match the inferred star formation history from Gaia \citep{Ruiz18} and to explain the recent abundance evolution (or lack thereof) in the solar neighborhood.
This picture is connected to the broader, currently accepted view of Milky Way formation. 
The Galaxy consists of a central bulge; a thick disk, which has a larger scale height and shorter radial extent; a thin disk, where stars have smaller vertical dispersions and ongoing star formation; and a halo made up of old stars from the Milky Way as well as accreted stars from a variety of dwarf galaxy mergers \citep{Fuhrmann98, McWilliam03, Sara}. 

Works tend to agree that the kinematically and spatially thick disk overlaps significantly with the stars formed early in the galaxy's history, where SNe~II had significantly enriched the gas in $\alpha$-elements, but SNe~Ia had not yet had sufficient time to add large numbers of iron-peak elements \citep{Tinsley79, Matteucci1986, Matteucci09, Kobayashi2011}.
These stars are therefore enhanced in $\alpha$-elements and are generally old, although debate is still ongoing about whether these stars started forming almost 14 Gyr ago, shortly after reionization, or whether there was a delay until about 10 Gyr ago, as well as whether there was a sharp burst of formation or a more continuous process \citep{Aguirre2018, Xiang2022}.
Around $\sim 8$--$11$~Gyr ago, the galaxy associated with the Gaia--Enceladus Sausage (GSE) merged with the Milky Way \citep{Belokurov2018, Myeong2018, Kim2021, Lane2023, Liu2024}.
This system is thought to have been roughly one-tenth the mass of the Milky Way at that time, and as such brought in a significant number of stars, now generally moving retrograde through the stellar halo. Furthermore, it potentially introduced a large amount of unprocessed gas \citep{Kordopatis2020} that altered the chemistry of stars formed after this time.
More recently formed stars are generally associated with the physically and kinematically thin disk.
These stars tend to have iron-rich and $\alpha$-poor compositions, and the combination of chemical and age information has suggested that star formation and chemical enrichment proceeded more rapidly in the inner Galaxy and more slowly in the outer galaxy \citep{Hayden15, Nidever24}. There is also strong evidence that significant numbers of stars migrate radially through the galaxy, changing their galactocentric radii and sometimes their height above the plane \citep{Schonrich09, Kubryk15, Sharma21}.
Today, we see that the Galactic bar, and the interactions between the bar and the disk, funnel stars and gas from the disk into the bulge \citep{Binney91, Bissantz02, Lopez05}. While it is challenging to work back in time, and bars in galaxies tend to be transient or short-lived features \citep{Tacchella15, Nelson16}, constraints on the Milky Way bar's formation epoch are emerging. The nuclear stellar disk provides an age tracer for the bar, with \citet{Sanders2024} finding that this structure contains significant populations older than $\sim8$~Gyr, suggesting the bar may have formed before or around the time of the GSE merger. These timing constraints provide important context for interpreting the chemical signatures of bar-driven gas flows. It is possible or even likely that these significant events in the broader Galaxy may have affected the star formation in the bulge, representing additional peaks in star formation at particular times and durations with particular compositions \citep{Elmegreen99, McWilliam03, Sara}.

Recent reanalyses of microlensed bulge dwarfs challenge this picture. In particular, \citet{Joyce2023} recompute ages for the \citet{Bensby17} microlensed sample using customized MESA Isochrones and Stellar Tracks (MIST)--based isochrones and find an age--metallicity distribution that is markedly older and narrower than previous findings \citep{Bensby17}. Furthermore, the field of Galactic archeology has advanced through large-scale spectroscopic surveys such as Gaia-ESO \citep{Gilmore12} and APOGEE \citep{Majewski17}, providing detailed abundance patterns that demand comprehensive galactic chemical evolution (GCE) models to interpret \citep{Ballero07, Grieco12, Molero2024, Dubay2025}.

\begin{figure*}[ht!]
\centering
\includegraphics[width=\linewidth]{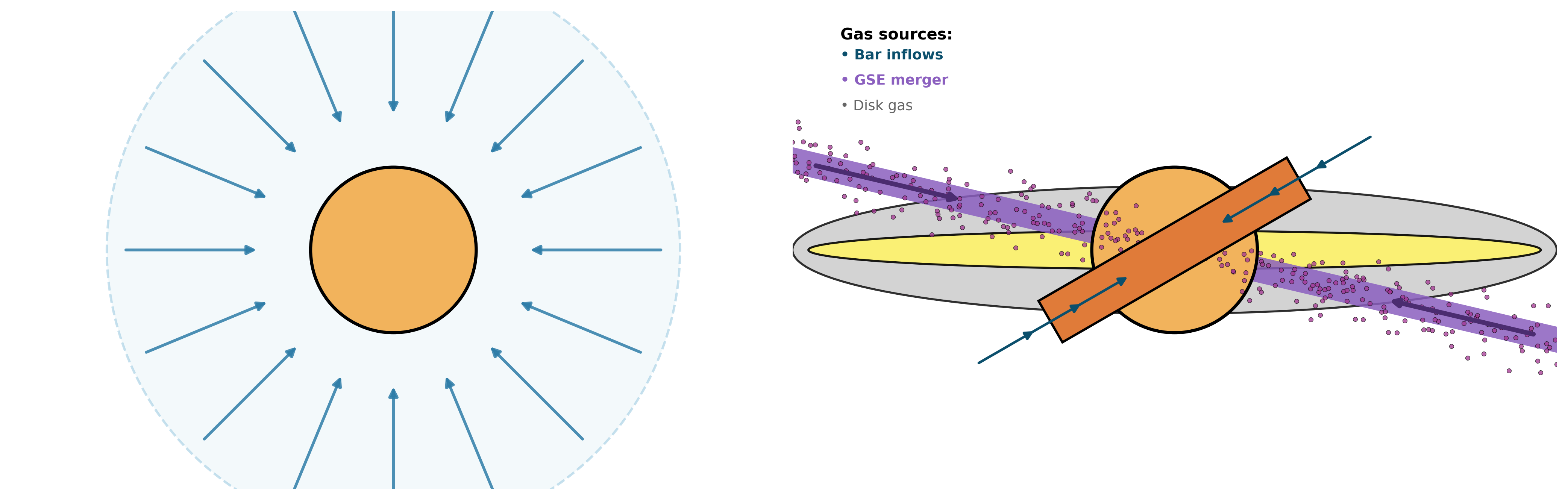}

\caption{
Schematic illustration of the two-infall model used in this work.
Left: face-on (looking down the Galaxy's rotation axis) view of the early bulge depicting the first infall (shown by the blue arrows), represented as a rapid, centrally directed collapse that builds the early bulge (orange circle) in the absence of a bar or extended disk.
Right: diagonal view of the later Galactic bulge during the second infall, where gas may be funneled inward along the existing bar (orange rectangle) and disk (the yellow ellipse represents a thin disk and the gray ellipse represents a thick disk), while additional material can arrive along more radial paths consistent with a GSE-like accretion event (the purple bar with red dots represents the GSE merger).
All of the channels supply fuel to the same central region but differ in geometry and angular momentum, motivating the distinct evolutionary signatures explored in this study.
}
\label{fig:intro-cartoon}
\end{figure*}

To guide the interpretation of our model assumptions, Figure~\ref{fig:intro-cartoon} summarizes the geometric and physical structure of the two-infall prescriptions adopted in this work. 
The left panel represents the initial rapid collapse phase that builds the early bulge, characterized by centrally concentrated gas inflows without a preexisting bar or extended disk. 
The right panel illustrates the subsequent infall episode (\textit{the second infall}), in which the established thin and thick disks together with the bar channel gas inward, while additional material may arrive along more radial trajectories consistent with a GSE-like accretion event. This bar-driven transport mechanism is supported by simulations (e.g., \citealt{Fanali2015}) and observations indicating that barred galaxies often exhibit flattened radial abundance and stellar population gradients alongside signs of central gas accumulation, consistent with efficient radial mixing \citep{Seidel2016, Fraser2020, Neumann2020}.
Each of the processes feed the same central reservoir but differ in geometry, timescale, and angular momentum.

Our primary goal is to obtain a coherent large-scale characterization of the bulge's chemical evolution; the resulting age--metallicity relations (AMRs) then emerge as a constrained by-product of this analysis rather than a standalone objective.  
We use GCE simulations to reproduce the observed bulge MDF, $\alpha$-element abundance patterns, and stellar age distributions.  
By demanding simultaneous agreement with these constraints, we can probe the relative contributions of a rapid ``classical'' formation phase (e.g., early dissipative collapse and mergers) and subsequent secular evolution driven by the Galactic bar (e.g., disk instabilities and bar buckling), as well as the contribution of major mergers such as GSE.

Although our models do not explicitly follow radial migration, the inferred chemo-chronological patterns can still be interpreted within the broader context of bar-driven evolution and the associated redistribution of stars and gas in the inner Galaxy.
This refines our picture of the Milky Way bulge and informs broader models of bulge formation in external galaxies, many of which show similarly multimodal MDFs and abundance gradients \citep{Kobayashi2020, Nidever24}.

\section{Observational Data}

\begin{figure}
    \centering
    \includegraphics[width=1.05\linewidth]{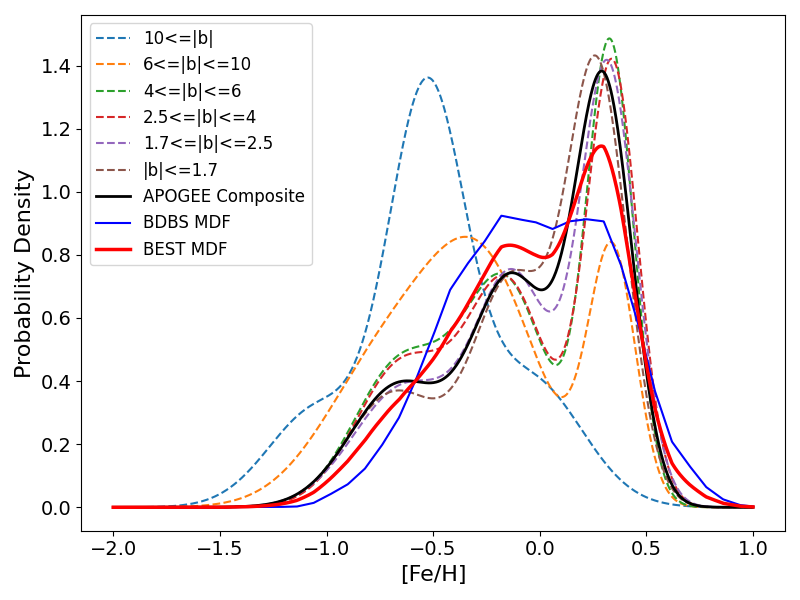}
\caption{
The figure shows the $\mathrm{[Fe/H]}$ distribution for the Milky Way bulge. The dashed lines indicate individual MDF fits from APOGEE DR16 across various Galactic latitude bands ``$|\mathrm{b}|$.'' The APOGEE composite (thick black line) is the latitude-weighted average of the APOGEE fits. The BDBS MDF (blue line) is derived from red clump stars \citep{Johnson2022}. The composite MDF (thick red line) represents the final, equally weighted composite observational target ($50\%$ APOGEE, $50\%$ BDBS) used for the GCE model optimization in this study. The target exhibits the characteristic bimodal distribution with peaks near $\mathrm{[Fe/H]} \sim -0.3$ and $\mathrm{[Fe/H]} \sim +0.3$.
}
    \label{fig:best_mdf}
\end{figure}

The primary optimization target is a composite MDF combining data derived from red clump stars via the Blanco DECam Bulge Survey \citep[BDBS;][]{Johnson2022} and a parameterization of the red giant branch (RGB)-based APOGEE DR16 MDFs at various latitudes provided by \citet{Rojas2020}. 
The BDBS component used here comprises $\sim$78,000 red clump stars from the \citet{Johnson2022} catalog at $|b| \approx 6^\circ$, while the APOGEE bulge sample comprises $\sim$20,000 RGB stars across multiple latitude bins. 
These sample sizes provide constraints on the MDF shape, though we note that the effective number of independent metallicity bins ($\sim$50--100) is considerably smaller than the raw star counts. 
The latter dataset is included primarily to improve coverage at the metal-poor ([Fe/H] $<$ $-1$) end, which is not well sampled by red clump stars in \citet{Johnson2022}.  

The BDBS metallicities are derived photometrically from DECam $u-i$ photometry and carry systematic uncertainties inherent to this approach; \citet{Johnson2022} report a typical metallicity uncertainty of $\sim0.2$~dex, and photometric metallicity estimates are known to be less precise than spectroscopic measurements. 

The two datasets were normalized by applying a latitude-based weighting of source counts based on the equation
\begin{equation}
\frac{N}{N_{o}}=1.029e^{0.476\cdot b}
\end{equation}
where $\frac{N}{N_{o}}$ represents the fraction of stars observed at a given Galactic latitude ($b$).  The scaling relation is based on a fit to the latitude-dependent angular mass density relation at $l$=0 provided by \citet{Zoccali2018}. We note that this relation is most accurate near $l=0$ and may lose accuracy at larger longitudes where the bar and peanut bulge dominate the mass distribution. The APOGEE fields used here span $|l| \lesssim 10^\circ$, while the BDBS coverage extends to similar longitudes with a concentration toward the minor axis. A more sophisticated treatment incorporating longitudinal dependence in the normalization may be warranted for future work, particularly given the bar's influence on stellar density gradients away from the minor axis.  The BDBS data were interpolated onto the APOGEE [Fe/H] grid, and the final composite MDF is shown in Figure~\ref{fig:best_mdf}.
We note that the BDBS distribution (derived from red clump stars) appears less visually bimodal than the APOGEE distribution (derived from RGB stars). This difference likely stems from the distinct selection functions and tracer sensitivities of the two surveys.
Notably, the two MDFs show significant disagreement at the very metal-rich ([Fe/H] $>$ +0.3) and very metal-poor ([Fe/H] $<$ $-1$) ends. The APOGEE MDF exhibits a more pronounced metal-rich tail and contains more low-$\alpha$ populations at high metallicity, while the BDBS distribution shows a stronger metal-poor component. These differences reflect the distinct stellar populations probed by each tracer (RGB versus red clump), their spatial coverage, and associated selection biases. Because our model is fit to the composite MDF, these systematic differences propagate into our results, particularly affecting constraints on the late-time enrichment history. Future work exploring the APOGEE-only MDF may help disentangle these effects.

\section{Methodology}

\subsection{GCE Modeling Machinery}

The One-zone Model for the Evolution of Galaxies and Abundances \citep[OMEGA;][]{Cote2016}\footnote{\url{https://github.com/NuGrid/NUPYCEE}} is a single-reservoir chemical evolution code that follows the time-dependent enrichment of a well-mixed galactic gas zone. 
In its baseline configuration, OMEGA assumes instantaneous mixing within a single gas reservoir and evolves the chemical composition under global prescriptions for star formation, stellar ejecta, and gas accretion. 
We incorporate nucleosynthetic yields through the SYGMA/NuPyCEE\footnote{\url{https://nugrid.github.io/NuPyCEE/}} code.

For massive stars ($M_{\star} \gtrsim 8\,M_{\odot}$), each GCE model adopts \emph{one} of several alternative massive-star-yield grids: the rotating models of \citet[hereafter LC18]{Limongi2018}, the nonrotating models of LC18, or the Nomoto/Chieffi SN-II-and-hypernova-enabled grids \citep{Chieffi2013, Nomoto2013}. 
These massive-star-yield sets are treated as distinct categorical ingredients in our model grid, and no model combines more than one massive-star-yield set.
Enrichment from low- and intermediate-mass stars ($M_{\star} < 8\,M_{\odot}$) is taken from the Monash yields of \citet{Karakas2010}.
SN\,Ia enrichment uses the metallicity-dependent yields of \citet{Shen2018} and \citet{Gronow2021}, coupled to several standard DTDs (power law, Gaussian, and exponential), following \citet{Greggio2005}, \citet{Strolger2004}, \citet{Matteucci2001}, and \citet{Maoz2014}. 
These SN\,Ia prescriptions constitute an additional categorical dimension in the model grid.
The initial chemical composition of gas---both at the start of the simulation and for any newly accreted inflow---is drawn from the STELLar ABundances (STELLAB) library provided by NuPyCEE \citep{Pignatari2016, Ritter2018, Battino2019}. This library consists of observed stellar abundance patterns (i.e., real interstellar medium (ISM)/stellar population abundances across elements) rather than pure primordial (H + He) composition. 
Thus, inflowing material is initialized with nonzero metallicity values representative of interstellar/circumgalactic gas, as inferred from observations.  
The initial metallicity is set by the categorical parameter [Fe/H]$_{\rm ini}$ (Table~\ref{tab:parameter_space}), which selects from the discrete STELLAB abundance patterns at [Fe/H] = 0.0, $-0.5$, $-1.0$, $-1.5$, and $-2.0$~dex. In our current implementation, this initial metallicity remains constant for all accreting gas throughout the simulation---i.e., both the primordial reservoir and subsequent infall episodes draw from the same STELLAB pattern. This simplification means that the metallicity of accreting gas does not evolve with time nor respond to prior enrichment in a circumgalactic medium (CGM). While this is a limitation, the posterior strongly prefers initial metallicities of [Fe/H] $\sim -0.5$ to $-1.0$~dex (Tables~\ref{tab:param_hdi_degeneracy} and \ref{tab:catagorical_params}), consistent with moderately preenriched infall. Future implementations could incorporate a time-evolving CGM metallicity that tracks the galaxy's enrichment history, which would be particularly important for interpreting the transition between infall episodes.

All chemical enrichment occurs in a single well-mixed bulge gas reservoir.  
Gas flows (inflows or outflows) simply add to or remove mass from this same reservoir, with instantaneous mixing and a homogeneous chemical composition throughout.

An extended version of OMEGA, OMEGA+ \citep{Cote2018, Cote2019}, was designed to model multiphase and multiepisode galactic evolution. 
OMEGA+ introduces a chemically evolving CGM that can exchange mass and metals with the central galaxy, enabling a two-zone mass-flow architecture rather than a strictly single-zone system. 
It also supports arbitrary user-defined gas inflow histories, allowing multiepisode accretion, such as the two-infall structure implemented in this study.

For this study, we implemented a further internal extension to OMEGA+, N'OMEGA+\footnote{\url{https://github.com/nialljmiller/MDF_GCE_SMC_DEMC}}, which permits time-dependent SFE, so that the efficiency may change across evolutionary phases. It also includes an SFE-aware time stepping to ensure the star formation rate is never undersampled.

In this study, we employ N'OMEGA+ in an inflows-only configuration.  
This choice is motivated by both physical and practical considerations.
First, the deep gravitational potential of a bulge makes strong large-scale galactic winds unlikely---SN feedback is unlikely to overcome the bulge's binding energy, so most of the gas would be retained and rapidly converted into stars \citep{Elmegreen99,Barbuy18}. 
Second, including outflows would introduce additional unconstrained parameters (e.g., mass-loading factor, wind-onset criteria, and wind metallicity), which would expand the parameter space and undermine the constraining power of our posterior analysis.  
Finally, observations of the bulge suggest a short, intense star formation history with rapid gas consumption and little ongoing gas loss that is consistent with an inflow-only gas-retention scenario \citep{Barbuy18,Matteucci2021}.

This configuration isolates the effects of early and late gas accretion in shaping the bulge's MDF bimodality and [$\alpha$/Fe] trends, while allowing a two-component formation history---an initial rapid collapse followed by a prolonged, slower accretion---motivated by theoretical bulge formation models \citep{Ballero07, Takuji2012, Molero2024} and observational indications of multiple stellar populations in the bulge \citep[e.g.,][]{Queiroz2021, Molero2024}.

The gas inflow rate follows a two-component exponential model that corresponds to accretion from a gas reservoir whose mass declines approximately exponentially in time.
This provides a minimal parameterization that captures an early rapid assembly followed by a gradual decline, while remaining directly comparable to previous bulge and disk studies \citep{Ballero07}. Given the quality and dimensionality of the available bulge constraints, the data do not yet justify adopting a functional form more complex than
\begin{equation}
\medmuskip2mu
\frac{d\sigma_{\mathrm{gas}}}{dt} =
\begin{cases}
A_1 \cdot e^{-t/\tau_1}, & t < t_{\mathrm{max},2}, \\
A_1 \cdot e^{-t/\tau_1} + A_2 \cdot e^{-(t - t_{\mathrm{max},2})/\tau_2}, & t \geq t_{\mathrm{max},2},
\end{cases}
\end{equation}
where $\sigma_{\mathrm{gas}}$ represents the gas surface density, $A_1$ and $A_2$ are normalization constants for the respective infall episodes, $\tau_1$ and $\tau_2$ denote the infall timescales, and $t_{\mathrm{max},2}$ determines the onset of the second episode. 
The first episode represents the early collapse producing the high-$\alpha$ stellar population, while the second infall accounts for delayed gas accretion that gives rise to the low-$\alpha$ population.

\subsection{Parameter-space Definition}

In order to better constrain this problem, we perform a wide parameter search using the resources of the University of Wyoming Advanced Research Computing Center (ARCC). 
We use their main cluster, Medicinebow, which hosts 25 dual-socket 48-core AMD-EPYC-9454 nodes.
This involves numerous trials using N'OMEGA+, with the goal of producing a posterior that allows us to constrain the physical conditions of the bulge formation process, especially the timing and duration of gas infall events.
The origins and justifications for each parameter choice are summarized in Table~\ref{tab:parameter_space}.

\begin{table*}[htbp!]
\centering
\begin{tabular}{lll}
        \hline
        Parameter & Range/Options & Source \\
        \hline
        $(\mathrm{Fe}/\mathrm{H})_{\rm ini}$ (initial composition) & +0.5 to  2.0 dex & [1,2,3] \\
        IMF prescription & Salpeter, Chabrier, Kroupa & [3,4,5] \\
        SN~Ia yield model & Gronow, Shen & [6,7,1,8] \\
        Stellar yields (massive + AGB) & Karakas AGB + LC18 with rotation, Nomoto HNe & [9,10,11,12] \\
        SN~Ia DTD & Power law, Gaussian, exponential & [13,14,15,16] \\
        $\sigma_2$ (second-to-first-infall strength ratio) & 0.1--10.0 & [17,3,18] \\
        $t_1$ (first-infall onset time) & 0.005--0.5 Gyr & [3,18] \\
        $t_2$ (second-infall onset time) & 0.1--10.0 Gyr & [3,19] \\
        $\tau_1$ (first-infall timescale) & 0.001--0.1 Gyr & [3,17] \\
        $\tau_2$ (second-infall timescale) & 0.1--10.0 Gyr & [3,17,20] \\
        SFE (first-phase SFE) & 0.1--27.0 Gyr$^{-1}$ & [21,22,23,3] \\
        $\Delta$SFE (multiplicative SFE drop at $t_2$) & 0.01--0.85 & [17,3,24] \\
        $M_{\max}$ (IMF upper-mass cutoff) & 60--130 M$_\odot$ & [25,26,27] \\
        $M_{\mathrm{Bulge}}$ (final stellar mass normalization) & $0.9$ $2.0 \times 10^{10}$ M$_\odot$ & [28,29,30] \\
        $N_{\rm Ia}/M_\odot$ (SN~Ia normalization per M$_\odot$) & $\mathbf{1.0 \times 10^{-6}}$--$\mathbf{1.5 \times 10^{-3}}$ & [13,15,1,31] \\
        \hline
    \end{tabular}

\caption{
Parameter Space Explored in the Genetic Algorithm Optimization of Our Two-infall Bulge Models.  
For each parameter, we list the range or discrete options sampled and the principal literature sources that motivate these choices.
The parameters are defined as follows:
$(\mathrm{Fe}/\mathrm{H})_{\rm ini}$ is the initial bulk iron abundance (in dex) of the gas reservoir;
``IMF prescription'' is the functional form of the stellar IMF;
``SN~Ia yield model'' specifies the nucleosynthetic yields adopted for SNe~Ia;
``Stellar yields (massive + AGB)'' are the yields assumed for massive stars and AGB stars;
``SN~Ia DTD'' gives the assumed functional form of the SN~Ia DTD;
$\sigma_2$ is the ratio of the total mass accreted in the second infall episode to that of the first;
$t_1$ and $t_2$ are the onset times of the first and second gas infall episodes, respectively (in Gyr after the start of the Universe);
$\tau_1$ and $\tau_2$ are the corresponding exponential infall timescales (in Gyr);
\mbox{SFE} is defined as the proportionality constant in 
$\mathrm{SFR} = \mathrm{SFE} \times M_{\rm gas}$, with units of Gyr$^{-1}$ (i.e., 
the fraction of the gas reservoir converted into stars per Gyr).
$\Delta$SFE is the multiplicative drop in SFE applied at $t_2$;
$M_{\max}$ is the upper-mass cutoff of the IMF (in M$_\odot$);
$M_{\mathrm{Bulge}}$ is the final stellar mass normalization of the bulge (in M$_\odot$);
and $N_{\rm Ia}/M_\odot$ is the normalization of the SN~Ia rate per unit stellar mass formed.
The numbered source labels in the final column correspond to:
[1] \citet{Kobayashi2020},
[2] \citet{Rojas2020},
[3] \citet{Salpeter1955},
[4] \citet{Chabrier2003},
[5] \citet{Kroupa2001},
[6] \citet{Gronow2021},
[7] \citet{Nomoto2018},
[8] \citet{Keegans2023},
[9] \citet{Limongi2018},
[10] \citet{Chieffi2013},
[11] \citet{Karakas2010},
[12] \citet{Cescutti2018},
[13] \citet{Greggio2005},
[14] \citet{Wiseman2021},
[15] \citet{Maoz2014},
[16] \citet{Strolger2005},
[17] \citet{Spitoni2019},
[18] \citet{Dubay2025},
[19] \citet{Lian2020},
[20] \citet{Chen22},
[21] \citet{Bigiel2008},
[22] \citet{Leroy2013},
[23] \citet{Kennicutt2021},
[24] \citet{Spitoni2020},
[25] \citet{Weidner2013},
[26] \citet{Schneider2018},
[27] \citet{Figer2005},
[28] \citet{Licquia2015},
[29] \citet{Valenti2016},
[30] \citet{Calamida2015},
[31] \citet{Trueman2025}.
}
\label{tab:parameter_space}
\end{table*}

Below, we outline the physical motivation for each parameter and the rationale behind the adopted ranges.
Some parameters are sampled in logarithmic space ($\sigma_2$, $t_1$, $t_2$, $\tau_1$, and $\tau_2$), because their physically plausible values span several orders of magnitude. These parameters control the timing and strength of the gas accretion episodes and, in turn, the overall pace of enrichment.
The parameter space is designed to span the dominant physical drivers of bulge chemical evolution. Each variable influences the model's ability to reproduce the observed MDF bimodality, the location of the [$\alpha$/Fe] knee, and the age--metallicity distribution.

The infall times, $t_1$ and $t_2$, represent the time in Gyr after the start of the Universe that the respective infall started.
$t_1$ and $\tau_1$ set the onset and duration of the first gas infall episode. Varying $t_1$ allows the model to explore different possible delays between the formation of the protobulge and the onset of efficient gas supply from the halo or early mergers. 
Very short values ($t_1 \sim 0.005 0.5$~Gyr; $\tau_1 \sim 0.001 0.1$~Gyr) produce the rapid initial enrichment required to form an old, $\alpha$-enhanced population. This behavior is consistent with the observed age distribution of the bulge.

The $t_2$ and $\tau_2$ control the timing and duration of the second infall episode. Delayed, extended, second-infall parameters ($t_2 \sim 0.1 10$~Gyr; $\tau_2 \sim 0.1 10$~Gyr) introduce dilution and allow the formation of a younger, more-metal-rich component. The range in $t_2$ includes values consistent with the epoch of the GSE merger and accommodates bar-driven inflows.
In models with rapid early collapse (short $\tau_1$) and efficient star formation, the [$\alpha$/Fe]-versus-[Fe/H] knee is pronounced; when later gas accretion is drawn out (long $\tau_2$), Fe-peak dilution is enhanced and the metal-rich tail of the MDF becomes more extended.
Recent age catalogs suggest that very short $\tau_1$ may be less realistic, motivating the exploration of slightly longer early timescales \citep{Molero2024, Dubay2025}.

The ratio of the mass accreted in the second infall relative to the first is given by $\sigma_2$.
Therefore, low values ($\sigma_2<1$) emphasize a classical early bulge formation, whereas high values ($\sigma_2>5$) imply substantial later accretion from disk inflows or merger debris.
The effect of the second-infall parameters also interacts strongly with $\tau_2$. A large $\sigma_2$ combined with a long $\tau_2$ flattens the high-metallicity tail of the MDF, whereas small $\sigma_2$ preserves a strongly peaked, old-population MDF.

The first-phase SFE sets the rate of gas consumption and enrichment during the initial collapse.
$\Delta$SFE is the multiplicative change in efficiency applied at $t_2$ (e.g., SFE$_2 = {\rm SFE} \times \Delta$SFE).
Importantly, for a given model run, the SFE remains constant throughout each phase and does not vary with individual time steps (as in some GCEs that implement density-dependent star formation); instead, a single multiplicative adjustment is applied at the transition time $t_2$.
High, early SFE produces the classical $\alpha$-plateau; modest changes at $t_2$ allow control over the metal-rich sequence without driving unrealistic late-time bursts.

The choices of the SN\,Ia model (\citealt{Gronow2021} and \citealt{Shen2018}), the SN\,Ia DTD (power law, Gaussian, exponential), and normalization ($N_{\rm Ia}/M_\odot$) control the timing and amount of Fe-peak enrichment. 

We note that our method assumes a time-independent SN\,Ia fraction; however, \citet{Mazzola2020} demonstrate that the binary fraction is strongly correlated with metallicity, implying the SN\,Ia rate should be higher at early times and decline toward the present. Incorporating a metallicity-dependent SN\,Ia normalization is left for future work.

These assumptions influence the position of the [$\alpha$/Fe] knee and the late-time slope of [Fe/H], and span the plausible observational range of bulge SN\,Ia rates.
Massive-star and asymptotic giant branch (AGB) yields are drawn from the LC18 grids with rotation, supplemented by Nomoto \citep{Nomoto2018} hypernova yields at low metallicity. 
Rotation affects light-element and s-process yields at the 0.1--0.2 dex level \citep{Pignatari2008}, and hypernova channels modify [$\alpha$/Fe] at early times. 
These choices ensure coverage of the dominant yield uncertainties relevant for bulge enrichment.
The initial mass function (IMF) prescription (either \citealt{Salpeter1955}, \citealt{Chabrier2003}, or \citealt{Kroupa2001}) and upper-mass cutoff ($M_{\max}$ = 60 130 M$\odot$) regulate the fraction of high-mass stars and thereby the strength of the $\alpha$-element production. 
Top-heavy IMFs enhance early SN II enrichment; larger $M_{\max}$ values allow rare, very massive stars to contribute to early chemical patterns.

$M_{\rm Bulge}$ ($0.9$ $2.0 \times 10^{10}$ M$_\odot$) sets the total amount of star formation required. In practice, it acts as a normalization on the integrated gas inflow, ensuring that the model reproduces the observed present-day stellar mass of the bulge.
Finally, the choice of the initial metallicity sets the baseline chemical state of the infalling gas. The initial metallicity range ($+0.5$ to $-2.0$ dex) allows the model to test scenarios from pristine early collapse to preenriched material. Lower initial [Fe/H] values with short $\tau_1$ help reproduce the observed metal-poor tail \citep{Ryde2025}.
The upper bound of [Fe/H]$_{\rm ini}$ = +0.5 dex may appear high, given that the high-$\alpha$ population peaks at $\sim-0.3$ dex; however, this extended range was included to allow the algorithm to explore whether preenriched gas from prior star formation (e.g., in a protobulge or early inner disk) could contribute to the observed MDF. As shown in Table~\ref{tab:catagorical_params}, the posterior strongly disfavors initial metallicities above solar, with the highest high-posterior-density (HPD) fractions concentrated at [Fe/H] = $-0.5$ to $-1.0$ dex.

We note that the explored range of SN~Ia normalizations ($N_{\rm Ia}/M_\odot \sim 10^{-6}$--$1.5\times10^{-3}$) extends well below canonical literature values ($\sim1.3\times10^{-3}$; \citealt{Maoz2014}). This extended range was adopted to allow the algorithm to explore parameter space without artificial constraints. The best-fit value of $N_{\rm Ia}/M_\odot \approx 5.8\times10^{-4}$---roughly a factor of 2 below canonical values---warrants discussion. This lower normalization may reflect: (i) degeneracies with other parameters that affect the Fe-peak enrichment (e.g., yield choices or the IMF); (ii) the specific DTD assumed; or (iii) physical differences in the bulge environment. We caution that this value should not be interpreted as a robust measurement of the intrinsic SN~Ia rate, but rather as an effective parameter within our model.

\subsubsection{Time Steps}

In models with low SFE, stars form gradually, distributing SN II ejecta over a much longer timescale. 
In this regime, enrichment proceeds smoothly, and each time step samples only a modest change in the $\alpha$-element mass fraction. 
The same coarse time step adequately resolves the evolution, producing numerically stable and physically continuous abundance tracks without spurious looping or overshooting.

Conversely, for models with high SFE, gas is rapidly converted into stars within the first $\sim$30 Myr. This rapid burst produces an intense but short-lived injection of $\alpha$-elements from core-collapse SNe (SNe~II). When the numerical time step is coarse (e.g., $\Delta t \approx 30.0$ Myr), it fails to resolve this sharp enrichment event, capturing only the averaged rise and fall of the $\alpha$-element yield. 
The resulting integration error manifests as abrupt discontinuities and nonphysical oscillations in the chemical evolution tracks.

To ensure that the model correctly resolves both the quiescent and bursty phases of enrichment, we implement a dynamic time-step allocation scheme. 
The total number of integration steps, $N_{\mathrm{tot}}$, is fixed for computational consistency across all runs, but the temporal distribution of these steps is adaptive. 
60\% of the time-step budget is assigned to regions surrounding the two gas infall episodes, where enrichment evolves most rapidly. The remaining 40\% is distributed across the intervening and late-time epochs, capturing the slow evolution dominated by SN~Ia and AGB yields.

Each high-resolution window extends $\sim3\tau_i$ (where $\tau_i$ is the infall timescale) beyond the onset time $t_i$ of each infall episode, corresponding to $\sim95\%$ of the exponential accretion. 
This ensures that the model resolves the steep rise and decay of gas inflow and star formation during both the early ($t_1, \tau_1$) and secondary ($t_2, \tau_2$) enrichment phases. 
The dynamic grid is constructed at runtime from the input parameters so that time-step refinement automatically tracks any changes to $t_i$ or $\tau_i$ during parameter optimization. 

For this study, $N_{\mathrm{tot}} = 300$ provides a temporal resolution of $\sim1$--$3$~Myr during the two infalls, which is sufficient to capture the prompt $\alpha$ enrichment even for models with ${\rm SFE} \gtrsim 15~\mathrm{Gyr^{-1}}$. 
This stabilizes high-SFE models while avoiding unnecessary computation at late times, maintaining a physically consistent time resolution across the full range of Galactic bulge scenarios tested in this study.

\subsection{Complexity of Parameter Space}

The GCE parameter space presents unique optimization challenges that render traditional methods inadequate. The parameter space exhibits: 
\\\noindent(i) mixed categorical and (pseudo-)continuous variables; 
\\\noindent(ii) highly nonlinear parameter--observation relationships due to complex stellar physics; 
\\\noindent(iii) multiple degenerate solutions producing nearly identical observational signatures; and 
\\\noindent(iv) discontinuous response regions where small parameter changes yield dramatically different outcomes. 
\\ \noindent These characteristics violate the smoothness assumptions underlying gradient-based methods and create severe convergence difficulties for Markov Chain Monte Carlo (MCMC) approaches.

While conventional MCMC methods are well suited to sampling posteriors for a fixed, continuous model, they become increasingly expensive when the target distribution is high-dimensional and contains multiple well-separated regions of high probability, especially in the presence of discrete or categorical choices. 
In our case, the parameter space combines (pseudo-)continuous parameters with categorical model switches (e.g., IMF choice, SN~Ia yield set, or DTD form). 
A fully Bayesian treatment of these discrete dimensions would require transdimensional schemes such as reversible-jump MCMC \citep{Green1995}, together with a careful convergence assessment for each effective model configuration, which is beyond the scope of this work.

We therefore use genetic algorithms (GAs) as a global optimization and exploration tool. 
GAs treat the objective function as a black box and do not rely on smoothness, gradients, or a particular covariance structure in parameter space \citep{Whitley1994}.
Their population-based updates allow the simultaneous exploration of multiple promising regions and help avoid trapping in local optima. 

However, standard GAs do not naturally produce calibrated posterior distributions, limiting their utility for uncertainty quantification. To address this, we implement a hybrid optimization machinery that combines the global exploration capabilities of GAs with Differential Evolution Markov Chain (DEMC) Monte Carlo moves for local refinement. This integration leverages the strength of GAs in navigating complex spaces while incorporating MCMC elements to enable probabilistic inference.

In addition to this hybrid approach, we perform independent SMC-DEMC MCMC analyses for each unique combination of categorical parameters (Section~\ref{sec:pure-mcmc}). These runs serve as a benchmark, allowing us to verify that the hybrid GA+DEMC exploration is consistent with the fully sampled posteriors for fixed model choices.

An alternative class of methods for navigating expensive, high-dimensional parameter spaces comes from the statistics and computer science literature \citep{Fang2005, Santner2013, Kleijnen2018}. 
These approaches, which include emulator-based inference using Gaussian Process regression and related surrogate modeling techniques, can in principle yield accurate posterior estimates from far fewer model evaluations than direct sampling, by interpolating over a sparse but well-designed grid of simulations \citep{Kaufman2011, Wibking2020}. 
Our current implementation already employs Latin hypercube sampling for initialization, which is a foundational element of this methodology. A natural extension of the present work would replace direct model evaluations with a trained emulator, potentially reducing the computational cost by 1 to 2 orders of magnitude while enabling more thorough exploration of the categorical dimensions.

\subsection{Hybrid GA with DEMC Integration}\label{sec:methods}
The hybrid GA, implemented using the Distributed Evolutionary Algorithms in Python \citep[DEAP;][]{DEAP}, initializes a population of $N_{\text{pop}} = 128$ individuals. Categorical parameters are randomly selected from discrete options with equal probability, while continuous parameters employ Latin hypercube sampling \citep{McKay1979} to ensure uniform coverage of the parameter space (Table~\ref{tab:parameter_space}).

Tournament selection with $k=3$ participants identifies the parents for subsequent generations. The selected parents undergo fitness-weighted crossover, where the inheritance probabilities for categorical parameters follow
\begin{equation}
P(\text{inherits}_j) = \frac{f_j^{-1}}{\sum_{k} f_k^{-1}}.
\end{equation}
$P$, the probability of categorical parameter inheritance, is limited to 0.75, to ensure genetic mixing. Continuous parameters combine through weighted averaging with stochastic perturbation:
\begin{equation}
x_{\text{child},i} = w_1 x_{\text{parent1},i} + w_2 x_{\text{parent2},i} + \mathcal{N}(0, \sigma_{\text{noise}})
\end{equation}
where the weights utilize fitness-based probabilities and $\sigma_{\text{noise}} = 0.05 \times |x_{\text{parent1},i} - x_{\text{parent2},i}|$.

Adaptive Gaussian mutation employs generation-dependent strength:
\begin{equation}
\sigma_{\text{mut}}(g) = \sigma_0 \times \left(1 - 0.75 \times \frac{g}{G_{\text{max}}}\right)
\end{equation}
with $\sigma_0 = 0.02$ and $G_{\text{max}} = 256$ generations. The categorical parameters mutate with 10\% probability to a random alternative.
To enhance exploration and prevent incomplete parameter-space coverage, we implement Voronoi-tessellation-based region identification.
Sparse regions, determined through cell area calculations in normalized parameter space, receive targeted exploration by the redirection of poorly performing individuals toward undersampled centroids. 
DEMC moves are integrated directly into each GA generation to provide local refinement and MCMC-like sampling. 
After population replacement, a configurable fraction (default: 40\%) of individuals undergoes one or more DEMC sweeps \citep{terBraak06}. Each selected walker proposes a new position based on the scaled difference of two randomly chosen peers plus Gaussian jitter:
\begin{equation}
\mathbf{x}' = \mathbf{x}_i + \gamma (\mathbf{x}_{r1} - \mathbf{x}_{r2}) + \boldsymbol{\epsilon}, \quad \boldsymbol{\epsilon} \sim \mathcal{N}(0, 10^{-9}),
\end{equation}
where $\gamma = 2.38 / \sqrt{2d}$ (with $d$ dimensions) by default, or $\gamma \approx 1$ every sixth generation for larger jumps. Proposals are reflected at bounds and accepted via the Metropolis--Hastings ratio using the same loss function as the GA fitness. This hybrid step improves local exploration while maintaining the GA's global search capability. The code to do this is publicly available at Zenodo via \dataset[doi:10.5281/zenodo.20418834]{https://doi.org/10.5281/zenodo.20418834} \citep{Miller2026zenodo} and GitHub (see footnote 10).
The algorithm first runs with the intention of high exploration (the first 32 generations), where the worst-performing 40\% of individuals are redirected to sparse Voronoi regions with a mutation probability of 0.8 and a crossover probability of 0.2. 
Subsequent convergence phases adjust to move only the least-well-performing 10\%, with adaptive mutation/crossover rates based on the fitness spread and mean nearest-neighbor distances in parameter space.

\subsection{Experiment Design}

Models optimize exclusively against the composite MDF shown in Figure~\ref{fig:best_mdf} using the ensemble loss function in Equation~(\ref{eq:ensemble_loss}), subject to a physical plausibility constraint. After each N'OMEGA+ model is evolved, the total baryonic mass at the final time step is computed as the sum of the locked stellar mass and the remaining gas mass. Models are retained only if this final mass lies within the physically motivated window $5\times10^{9}\,M_\odot < M_{\rm final} < 3\times10^{10}\,M_\odot$.

This allows us to identify viable two-infall scenarios and to assess which AMR---that of \citet{Bensby17}, that of \citet{Joyce2023}, or neither---arises naturally from models that reproduce the present-day metallicity distribution. The AMR is not used as an optimization target; instead, we compare the postoptimization synthetic AMRs against both observational relations to determine which one is implied by the physics required to match the MDF. We note that incorporating the AMR as a simultaneous fitting constraint alongside the MDF is a natural extension of this work and will be explored in a forthcoming study.

The fitness evaluation uses an ensemble loss function composed of three complementary terms:
\begin{equation}\label{eq:ensemble_loss}
L_{\rm ensemble} = 0.7\,L_{\rm WRMSE} + 0.2\,L_{\rm cosine} + 0.1\,L_{\rm Huber}.
\end{equation}
Weighted RMSE provides the primary sensitivity to the overall shape of the MDF.  
Cosine similarity measures how well the model reproduces the relative pattern of the distribution, independent of its absolute normalization.  
Huber loss ($\delta = 1$) reduces the influence of isolated bins where observational uncertainties or sampling noise produce large deviations.

\subsection{MCMC Analysis}
\label{sec:pure-mcmc}

In addition to the hybrid GA DEMC, we perform a set of complementary MCMC analyses to provide a trusted reference for our inferred posteriors. 
For each of the 541 unique combinations of categorical parameters (IMF prescription, SN~Ia yield model, and DTD form, etc.), we fix those categorical choices and run an independent DEMC sampler on the continuous parameters only. 
In practice, this means we treat the categorical parameters as a finite grid. 
For every point on that grid, we run a separate MCMC chain in the continuous subspace, rather than attempting a single sampler that jumps between different model choices. 

Each MCMC run starts from the same uniform priors over the continuous parameters used in the hybrid analysis (Table~\ref{tab:parameter_space}) and follows the DEMC annealing schedule described in Section~\ref{sec:methods} (i.e., a sequence of tempered intermediate distributions that gradually approach the target posterior). 
Because the categorical parameters are held fixed in each run, these analyses yield clean posterior samples for a single well-defined model configuration and provide baseline convergence diagnostics.

This exhaustive set of runs requires 541 separate MCMC experiments and serves as a benchmark for our GA+DEMC method. 
We can check that the regions of high posterior probability identified by the hybrid GA+DEMC procedure are consistent with those obtained from conventional MCMC for the same fixed categorical choices.

\section{Posterior Analysis}

\begin{figure*}[htb]
\includegraphics[width=\linewidth]{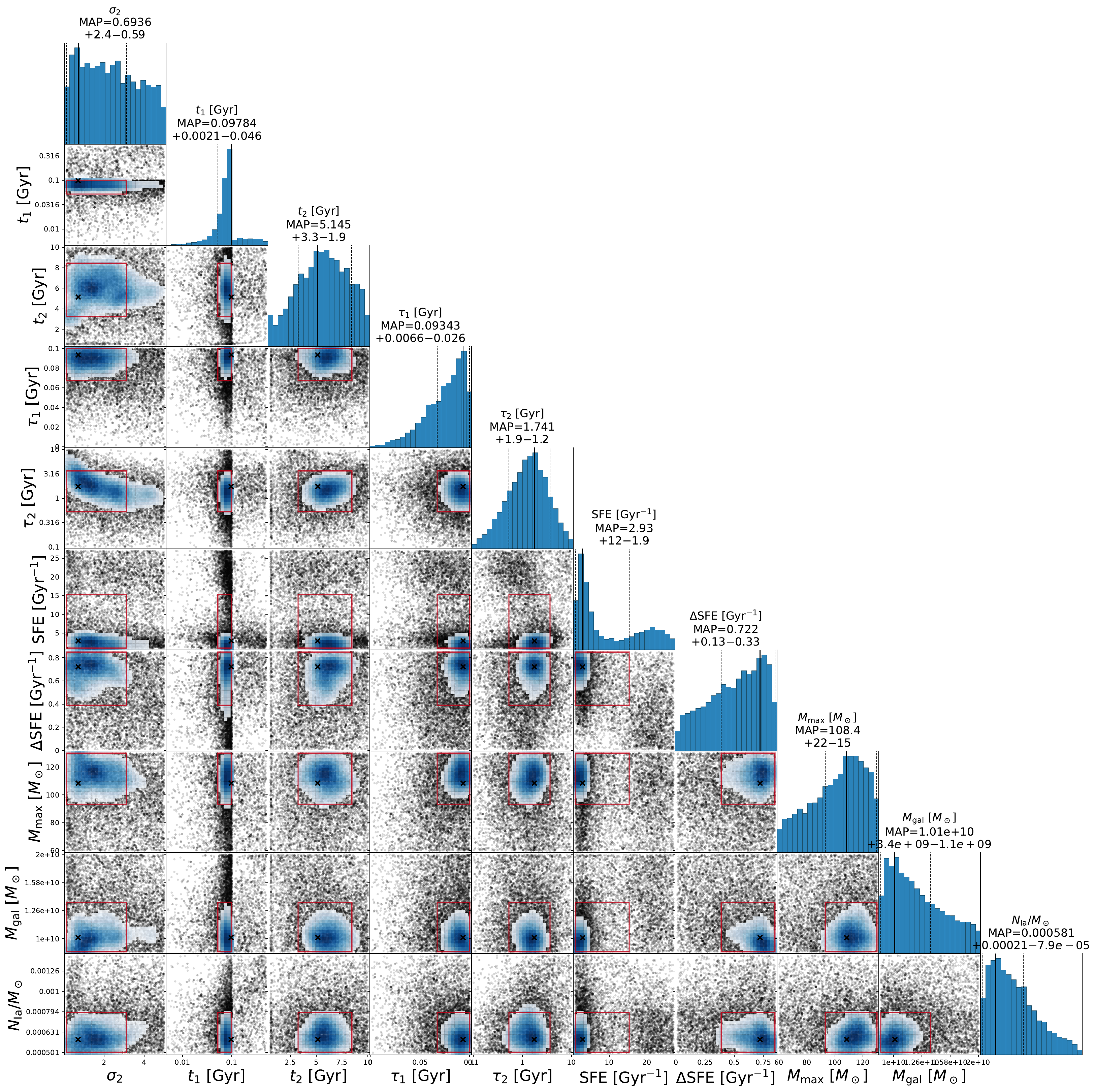}
\caption{Corner plot of posterior showing 1D histograms (diagonals) and 2D density (off-diagonals) continuous parameter groups. The black cross indicates the MAP. The red square highlights the HDI.}
\label{fig:param_correlations}
\end{figure*}

The GA was executed 16 times using identical hyperparameters and prior boundaries, each starting from a different random seed. 
All runs converged toward the same region of parameter space, demonstrating that the identified solution is robust and not an artifact of stochastic sampling. 
The combined catalog of evaluated models was then converted into an approximate posterior---a pseudo-posterior---by assigning weights based on the MDF loss. 
Figure~\ref{fig:param_correlations} shows the resulting pseudo-posterior distributions, which are far from Gaussian and reveal a generally broad and degenerate solution space. 
Several parameters remain only weakly constrained by the MDF. 
To quantify these constraints, we extract the maximum \textit{a posteriori} (MAP) values and 68\% highest density intervals (HDIs) for all continuous parameters.
The mass ratio of the two infalls is not well constrained, with MAP $\sigma_2 \approx 0.69$ and HDI $0.11$--$3.1$.
The first-infall time is notably more constrained than the second, where $t_1 \approx 0.1$~Gyr and HDI $0.09$--$0.1$~Gyr, while $t_2 \approx 5.15$~Gyr and HDI $3.25$--$8.45$~Gyr.
The first-infall timescale ($\tau_1$) is confined to a narrow range $\approx 0.09$~Gyr (HDI $0.07$--$0.10$~Gyr). 
The second-infall timescale is less constrained, with $\tau_2 \approx 1.7$~Gyr and HDI $0.5$--$3.7$~Gyr.
The SFE shows a bimodal distribution, with a more prominent ``high'' SFE distribution (SFE $\approx 3$~Gyr$^{-1}$) and a less prominent ``extremely high'' SFE distribution at $\approx 25$~Gyr$^{-1}$. 
Neither value is atypical when compared to previous GCE studies \citep{Matteucci09, Grieco12}.
This leads to a potentially misleading HDI of $1$--$15$~Gyr$^{-1}$. 
The SN~Ia normalization ($N_{\rm Ia}/M_\odot$) is also moderately well constrained ($\approx 5.8 \times 10^{-4}$, HDI $5.0 \times 10^{-4}$--$7.9 \times 10^{-4}$).

Table~\ref{tab:param_hdi_degeneracy} lists the MAP estimates and 68\% HDI for all continuous parameters, together with the strongest pairwise degeneracies.
The most pronounced linear correlation is between the infall times, timescales, and SFE, with a Pearson correlation ($\rho_w$) $\simeq -0.14$ for the $\tau_1$ SFE and $\rho_w \simeq -0.20$ for the $t_2$ SFE (the latter being the strongest correlation in the matrix). 
Parameters such as the change in SFE ($\Delta\mathrm{SFE}$), the IMF upper mass ($M_{\max}$), the bulge stellar mass ($M_{\mathrm{Bulge}}$), and the SN~Ia normalization ($N_{\rm Ia}/M_\odot$) exhibit only modest linear correlations but large mutual information (MI) with $\sigma_2$, $\tau_1$, $\tau_2$, and $t_2$.

\begin{table*}[ht]
    \centering
    \begin{tabular}{l c c c l l }
        \hline
        Parameter &
        MAP &
        HDI$_{\rm lo}$ &
        HDI$_{\rm hi}$ &
        Top-three $\rho_w$ &
        Top-three MI \\
        \hline
        $\sigma_2$ & 0.69 & 0.11 & 3.12 & 0.19 ($M_{\mathrm{Bulge}}$) & 0.85 ($M_{\max}$) \\
         &  &  &  &  -0.18 ($\tau_2$) &  0.84 ($t_2$) \\
         &  &  &  &  -0.07  ($t_2$)   & 0.81 ($\tau_1$) \\
        $t_1$ & 0.098 & 0.0521 & 0.10 & -0.17 ($M_{\mathrm{Bulge}}$) & 0.77 ($\sigma_2$) \\
         &  &  &  & 0.13 ($\Delta\mathrm{SFE}$) & 0.77 ($\Delta\mathrm{SFE}$) \\
         &  &  &  & 0.08 ($t_2$) & 0.76 ($N_{\rm Ia}/M_\odot$) \\
        $t_2$ & 5.15 & 3.24 & 8.44 & -0.20 ($\mathrm{SFE}$) & 0.84 ($\sigma_2$) \\
         &  &  &  & 0.17 ($\tau_2$) & 0.83 ($M_{\max}$) \\
         &  &  &  & 0.10 ($M_{\max}$) & 0.81 ($\tau_2$) \\
        $\tau_1$ & 0.093 & 0.067 & 0.10 & -0.19 ($\tau_2$) & 0.82 ($M_{\max}$) \\
         &  &  &  & 0.15 ($M_{\max}$) & 0.81 ($\sigma_2$) \\
         &  &  &  & -0.14 ($\mathrm{SFE}$) & 0.76 ($\tau_2$) \\
        $\tau_2$ & 1.74 & 0.53 & 3.65 & -0.19 ($\tau_1$) & 0.81 ($t_2$) \\
         &  &  &  & -0.18 ($\sigma_2$) & 0.80 ($\sigma_2$) \\
         &  &  &  & 0.17 ($t_2$) & 0.77 ($M_{\mathrm{Bulge}}$) \\
        SFE & 2.93 & 1.00 & 15.3 & -0.20 ($t_2$) & 0.44 ($\tau_1$) \\
         &  &  &  & -0.14 ($\tau_1$) & 0.42 ($t_1$) \\
         &  &  &  & 0.12 ($M_{\mathrm{Bulge}}$) & 0.40 ($\Delta\mathrm{SFE}$) \\
        $\Delta$SFE & 0.72 & 0.39 & 0.85 & 0.14 ($\tau_1$) & 0.83 ($M_{\max}$) \\
         &  &  &  & -0.14 ($M_{\max}$) & 0.80 ($\sigma_2$) \\
         &  &  &  & 0.13 ($t_1$) & 0.79 ($M_{\mathrm{Bulge}}$) \\
        $M_{\max}$ & 108.4 & 93.2 & 130.0 & 0.15 ($\tau_1$) & 0.85 ($\sigma_2$) \\
         &  &  &  & -0.14 ($\Delta\mathrm{SFE}$) & 0.84 ($M_{\mathrm{Bulge}}$) \\
         &  &  &  & 0.10 ($t_2$) & 0.83 ($t_2$) \\
        $M_{\mathrm{Bulge}}$ & $1.01\times 10^{10}$ & $9.00\times 10^{09}$ & $1.35\times 10^{10}$ & 0.19 ($\sigma_2$) & 0.84 ($M_{\max}$) \\
         &  &  &  & 0.18 ($N_{\rm Ia}/M_\odot$) & 0.79 ($\Delta\mathrm{SFE}$) \\
         &  &  &  & -0.17 ($t_1$) & 0.77 ($\tau_2$) \\
        $N_{\rm Ia}/M_\odot$ & $5.81\times 10^{-04}$ & $5.02\times 10^{-04}$ & $7.88\times 10^{-04}$ & 0.18 ($M_{\mathrm{Bulge}}$) & 0.81 ($\sigma_2$) \\
         &  &  &  & 0.05 ($\sigma_2$) & 0.77 ($\tau_2$) \\
         &  &  &  & 0.04 ($\mathrm{SFE}$) & 0.76 ($t_1$) \\
        \hline
    \end{tabular}
        \caption{Parameter Estimates from MAP and 68\% HDI, along with Top-three Pairwise Degeneracies Ranked by Weighted Pearson Correlation ($\rho_w$) and MI.}
    \label{tab:param_hdi_degeneracy}
\end{table*}

The pseudo-posterior exhibits several notable coupling patterns, which we describe below.
The infall mass ratio, $\sigma_2$, shows a mild negative correlation with the second-infall timescale $\tau_2$ ($\rho_w \simeq -0.18$), and both $\sigma_2$ and $\tau_2$ exhibit substantial MI with SFE (MI$_w \gtrsim 0.7$; Table~\ref{tab:param_hdi_degeneracy}). This indicates that the late-accreted mass fraction, its delivery timescale, and the global efficiency of star formation adjust together to preserve a viable MDF. A similar negative correlation is seen between the onset time of the second infall ($t_2$) and SFE ($\rho_w \simeq -0.20$), while $\tau_2$ is negatively correlated with the first-infall timescale $\tau_1$ ($\rho_w \simeq -0.19$). 
Together, these relationships produce the elongated ridges in the corner plot, corresponding to families of models in which changes to coupled parameters compensate for one another, leaving the final MDF and $\alpha$-element trends nearly unchanged.

These relationships reflect underlying physical trade-offs in bulge evolution.
Shifting gas to arrive later or over a longer timescale can be offset by adjusting the efficiency with which it is turned into stars and by redistributing mass between the first and second infall, so that the integrated enrichment history remains compatible with the present-day MDF data. 
Within this credible region, the effective enrichment timescale ($\tau_{\rm enrich} \sim \tau_{\rm infall}/\nu$) remains approximately constant at $\sim 0.03$--$0.1$~Gyr, ensuring the rapid early enrichment required by the $\alpha$-knee position and the old ages of metal-poor bulge stars. 

The negative correlation between $\sigma_2$ and $\tau_2$, together with their strong dependence on SFE (as quantified by the MI), encodes a late-time trade-off: a more massive or more concentrated second infall (higher $\sigma_2$, shorter $\tau_2$) must be compensated for by a lower SFE or a shift in timing to avoid overproducing young metal-rich stars.

In this sense, the coupled $\sigma_2$, $\tau_2$, SFE and $\tau_1$, $\tau_2$ degeneracies jointly control how aggressively the metal-rich peak of the MDF and the young end of the AMR can be built without violating the $\alpha$-element and bulge-mass constraints. 
Despite these broad individual constraints, the pseudo-posterior occupies a physically plausible region of parameter space that consistently reproduces both the observed MDF and the \citet{Joyce2023} AMR. 

Taken together, the pseudo-posterior favors a coherent evolutionary scenario where the preferred solution describes a rapid initial collapse phase, with a first infall peaking at $t_1 \approx 0.1$~Gyr, with a very short duration $\tau_1 \approx 0.1$~Gyr and an SFE of order a few~Gyr$^{-1}$. 
This is followed by a delayed lower-mass second infall episode, with onset at $t_2 \approx 5$~Gyr, corresponding to $\sim 9$~Gyr ago, with $\tau_2 \sim 1.7$~Gyr and $\sigma_2 \approx 0.69$.
This configuration naturally produces an early burst of $\alpha$-enhanced metal-poor stars, followed by a prolonged phase of diluted star formation that builds the metal-rich younger component---exactly the hybrid classical-plus-secular scenario favored by recent bulge studies \citep{Ballero07, Takuji2012, Molero2024}.

\begin{figure*}[htb!]
\centering
\includegraphics[width=0.45\linewidth]{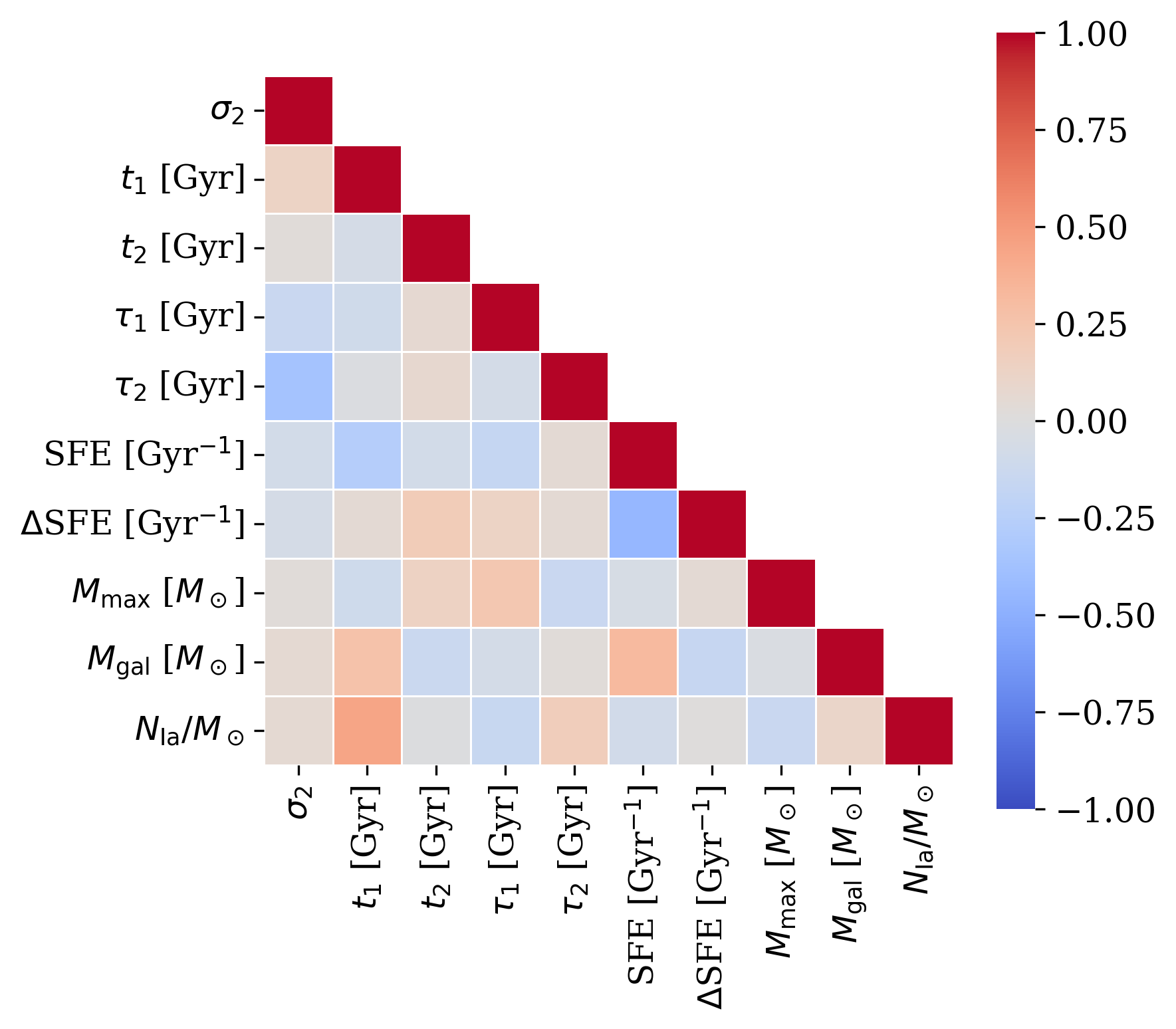}
\includegraphics[width=0.45\linewidth]{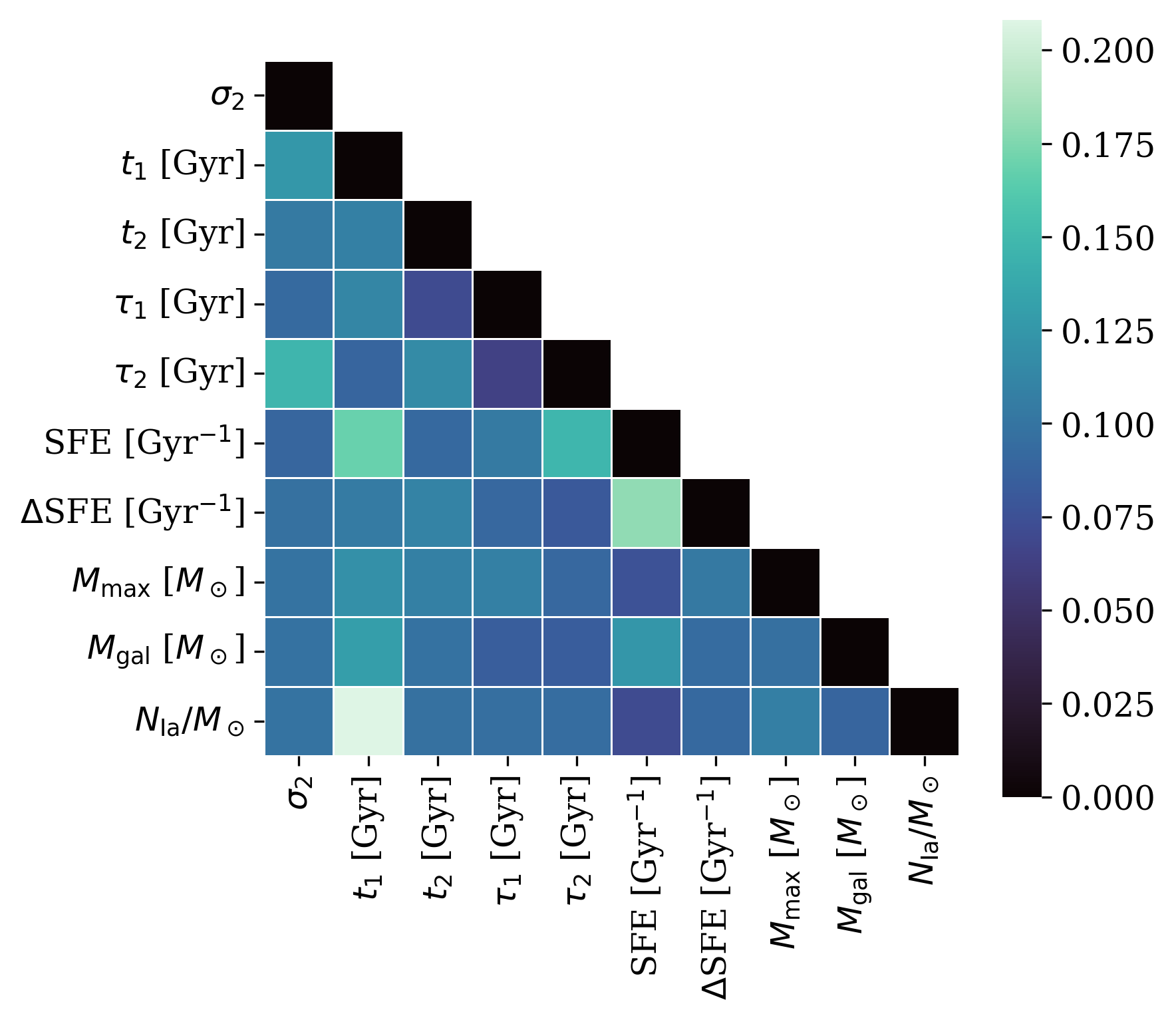}
\caption{Parameter dependency analysis.
Left: fitness-weighted Pearson correlation matrix.
Right: MI matrix that highlights both linear and nonlinear dependencies. 
Parameter pairs with high MI but low correlation indicate strong nonlinear coupling.}
\label{fig:correlations}
\end{figure*}

To visualize these relationships, we compute both the weighted Pearson correlation (linear relationships among well-fitting models) and the MI (linear and nonlinear dependencies).
Figure~\ref{fig:correlations} shows a matrix for each of these values.
The strongest correlations from the weighted Pearson matrix include a negative association between the onset time of the second infall ($t_2$) and the SFE with $\rho_w \approx -0.20$, indicating a trade-off whereby earlier second episodes require lower SFE to avoid over-producing metal-rich stars and to preserve the observed MDF shape.

Within the second-infall sector, $\sigma_2$ and SFE show only a weak linear correlation ($\rho_w \approx -0.05$) but a nonnegligible MI (MI$_w \approx 0.40$), highlighting that viable models balance the late-accreted mass fraction against the global efficiency of star formation to keep both the bulge mass and MDF in agreement with the data.
Parameter pairs with high MI but low weighted correlation include $\sigma_2$ with $M_{\max}$ ($\rho_w \approx 0.00$, MI$_w \approx 0.85$) and $M_{\max}$ with $M_{\mathrm{Bulge}}$ ($\rho_w \approx 0.09$, MI$_w \approx 0.84$).
These indicate nonlinear dependencies that potentially arise from threshold effects in enrichment, where changes in the IMF upper mass or the late-infall mass ratio sharply modify the contribution of very massive stars at fixed bulge mass and hence the detailed chemical pattern.

\subsubsection{Principal Component Analysis}
\begin{figure*}[ht!]
\centering
\includegraphics[width=0.48\linewidth]{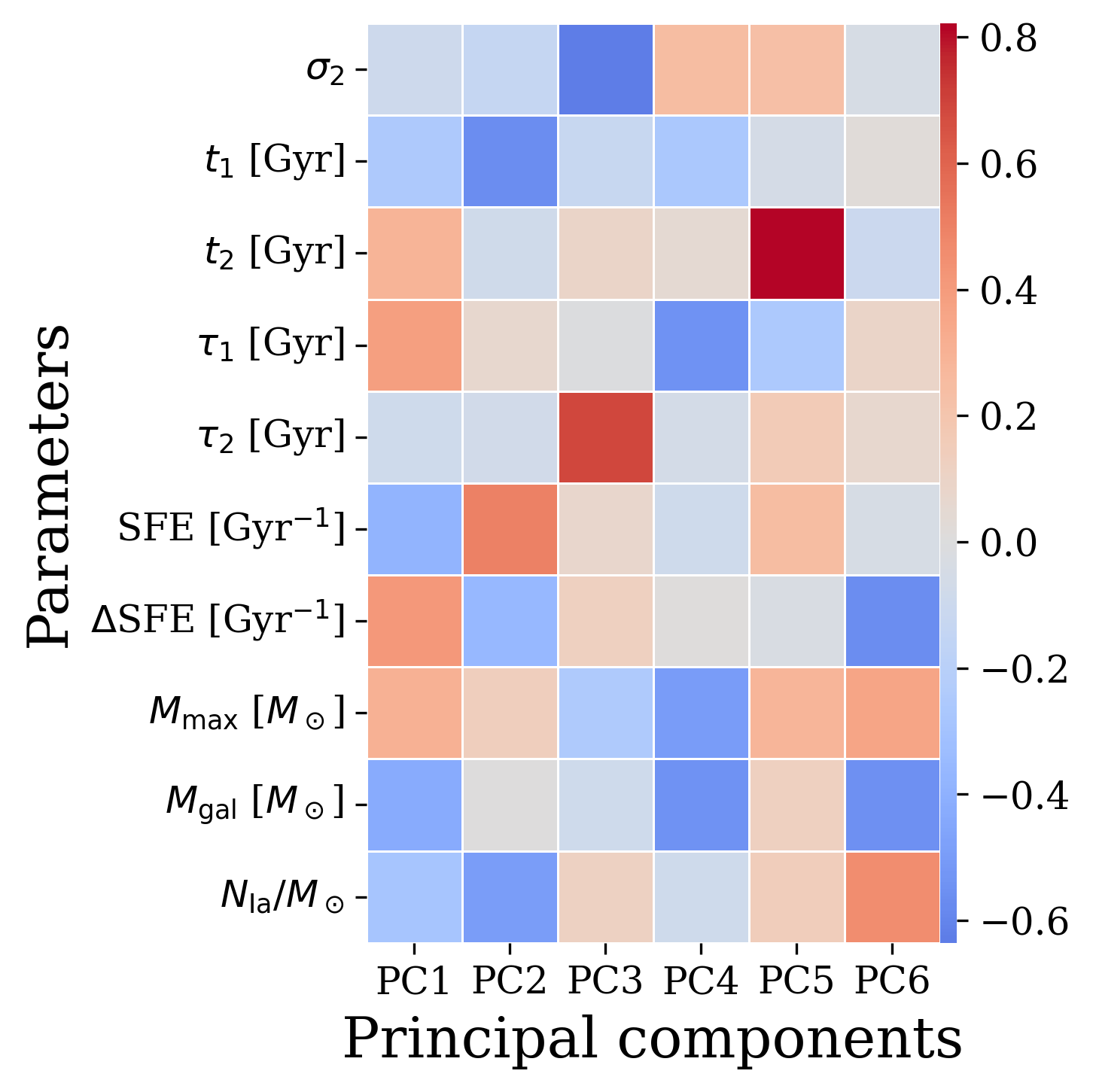}
\includegraphics[width=0.48\linewidth]{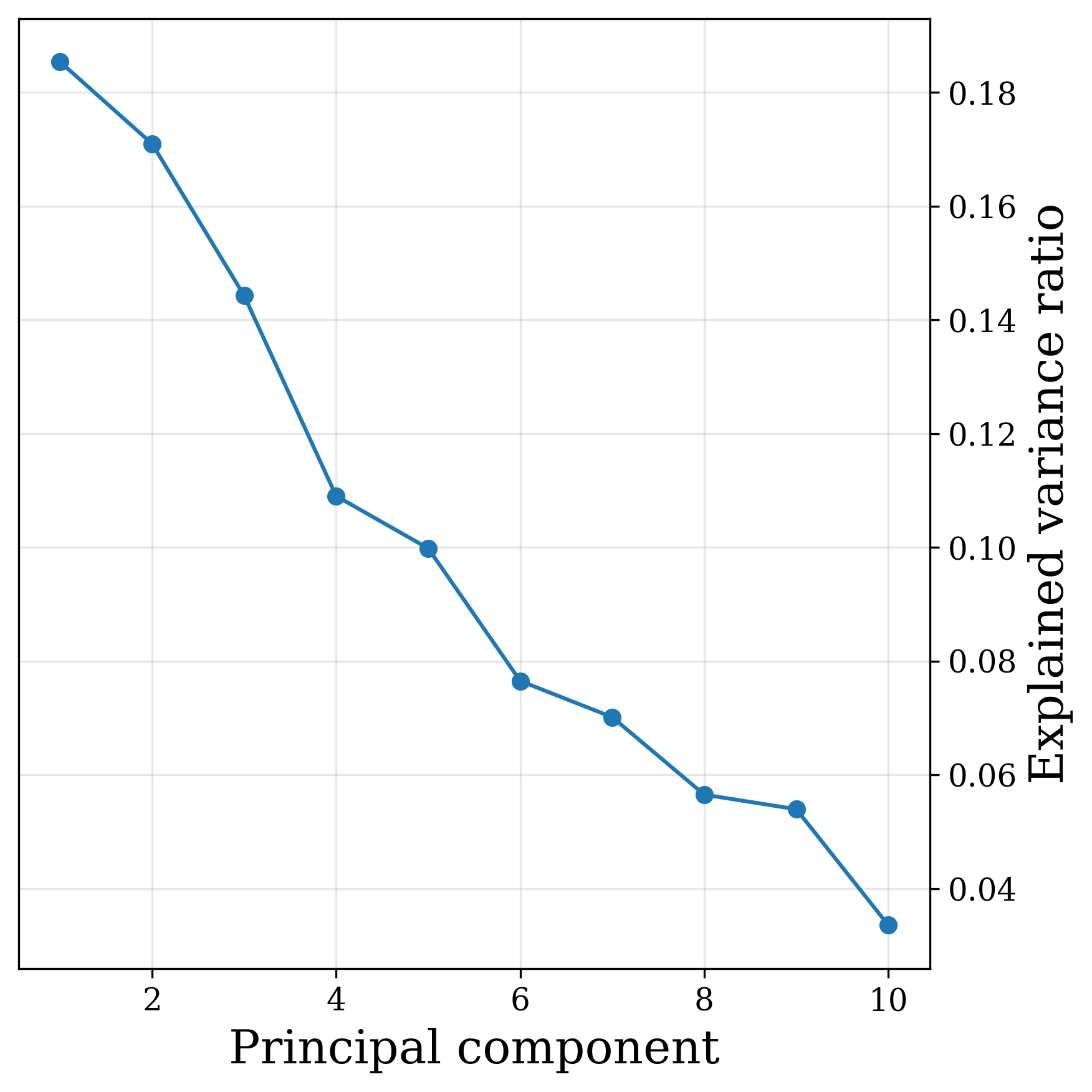}
\caption{PCA of the top-performing models.
Left: parameter loadings on the first six PCs,
showing which parameters contribute most strongly to each mode of variation.
Right: the variance explained by each PC. The first six PCs capture approximately 77.5\% of the total variance,
indicating substantial degeneracy in the parameter space.}
\label{fig:pca_analysis}
\end{figure*}

Principal component analysis (PCA) of the top-performing models demonstrates the fundamental degeneracy seen in the GCE parameter space.
Figure~\ref{fig:pca_analysis} summarizes the PCA of the top-performing models, showing both parameter loadings and the variance captured by each component.
The first PC (PC1) explains 21.3\% of the variance, with the strongest loadings on SFE ($-0.523$), $\tau_2$ ($+0.414$), $t_1$ ($-0.402$), and $\sigma_2$ ($-0.378$). 
This component represents a trade-off between SFE and the second-infall process, where the reduced SFE and earlier first-infall timing ($t_1$) balance the increased late-infall amplitude ($\tau_2$) and lower mass ratio between infalls ($\sigma_2$), maintaining the overall enrichment rates in the two-infall model.
The second PC (PC2) accounts for 14.9\% of the variance, dominated by loadings on $\tau_1$ ($-0.596$), $\sigma_2$ ($+0.487$), $M_{\max}$ ($-0.336$), and $\Delta\mathrm{SFE}$ ($+0.336$). 
This component shows the relationship between the first-infall strength and mass partitioning ($\sigma_2$), with adjustments in the IMF upper mass and SFE variation. PC2 differs from PC1 by focusing on early-phase gas supply and yield scaling rather than temporal trade-offs.
The third PC (PC3) explains 12.4\% of the variance, with key loadings on $t_2$ (+0.635), $M_{\max}$ (+0.374), $\Delta\mathrm{SFE}$ (+0.309), and $\tau_1$ (+0.301). 
This mode interprets the fine-tuning of the late-phase evolution, where extended second-infall timing ($t_2$) interacts with IMF limits and SFE changes to refine enrichment; the cumulative variance of PC1--PC3 at 48.6\% indicates a moderate reduction in dimensionality, with about half the variation captured by these primary modes, suggesting underlying physical constraints but persistent degeneracies in higher components.
The effective dimensionality, calculated using the participation ratio $d_{\text{eff}} = \frac{\left(\sum_{i=1}^{N} \lambda_i\right)^2}{\sum_{i=1}^{N} \lambda_i^2} \approx 8.0$, implies approximately eight independent constraints relative to the 10 nominal parameters. This indicates that observational data reduce but do not fully resolve the parameter-space complexity.
The eigenvalue spectrum shows a gradual decline without a sharp elbow, implying that variance is distributed across many components rather than concentrated in a few dominant modes. 
Directions with low eigenvalues (e.g., PC9 and PC10, explaining $\sim$5\% and 4\%, respectively) correspond to weakly constrained parameter combinations, such as those involving SN~Ia rates or galactic mass, which have limited impact on the fitted MDF and AMR, due to the one-zone approximation's averaging effects \citep{Cote2016}.

\subsubsection{Categorical Parameters}
\begin{table*}[]
    \centering
    \begin{tabular}{llc}
        Parameter & Category & HPD Fraction \\
        \hline
        Component array 
            & iniab output feh p000  & 0.0478 \\
            & iniab output feh m050  & 0.3841 \\
            & iniab output feh m100  & 0.3715 \\
            & iniab output feh m150  & 0.1048 \\
            & iniab output feh m200  & 0.1054 \\
        IMF 
            & Salpeter               & 0.0209 \\
            & Chabrier               & 0.5837 \\
            & Kroupa                 & 0.3954 \\
        SN\,Ia assumptions 
            & SN\,Ia Gronow          & 0.4507 \\
            & SN\,Ia Shen            & 0.5493 \\
        Stellar yield assumptions 
            & K10 LC18 R150          & 0.2245 \\
            & K10 LC18 Ravg          & 0.4755 \\
            & K10 LC18 R000          & 0.2830 \\
            & C15 N13 0 0 HNe        & 0.0170 \\
        SN\,Ia rate models 
            & Power law           & 0.4845 \\
            & Gaussian            & 0.4274 \\
            & Exponential         & 0.0881 \\
    \end{tabular}

\caption{Posterior HPD Fractions for All Discrete (Categorical) Model Ingredients---component abundance array (Where the files p000, m050, m100, m150, and m200 correspond to fixed initial metallicities of [Fe/H] = 0.0,  0.5,  1.0,  1.5, and  2.0 dex, respectively, with the associated $\alpha$-enhancement pattern inherited from the NuGrid abundance library), IMF, SN\,Ia Implementation, Stellar Yield Set, and SN\,Ia Rate (Delay-time) Model---within the HPD Subset of Models Shown in Figure~\ref{fig:big-corner}. 
None of the options in any category are driven to zero or unity.}    
    \label{tab:catagorical_params}
\end{table*}

Table~\ref{tab:catagorical_params} shows the posterior distribution of each categorical choice within the HPD-selected sample, the same HPD region quoted in Figure~\ref{fig:big-corner}.  
Within each group, the fractions are relatively broad but none of the options are driven to zero or unity, indicating that the MDF, AMR, and $\alpha$-element constraints do not strongly single out a unique component array, IMF family, SN\,Ia prescription, yield grid, or delay-time model.  
The IMF set is split roughly 60/40 between Chabrier and Kroupa, with a small Salpeter contribution, and the yield and SN\,Ia choices all retain substantial support.  
Changes in these discrete ingredients are absorbed by shifts in continuous parameters such as SFE, $M_{\max}$, and $n_b$, so the main constraints in this work apply to global enrichment timescales and mass budgets rather than to any specific categorical prescription.

\subsection{Model Limitations}
While our two-infall GCE model provides a method for constraining the bulge's enrichment history, several simplifying assumptions limit its scope and warrant discussion. 
First, we impose a two-infall functional form.  
The posterior indicates that more than one episode of gas supply is required to reproduce the MDF and abundance trends, but it does not demonstrate that there were exactly two.  
Any additional minor or closely spaced inflows---whether merger-driven, bar-mediated, or associated with secular disk processes---would be absorbed into the effective second episode in our one-zone parameterization, and therefore cannot be uniquely identified within this model.

The one-zone approximation assumes instantaneous and homogeneous mixing within a single well-mixed reservoir, which cannot capture the observed vertical metallicity gradients spanning $\Delta$[Fe/H] $\sim$ 0.3 dex across the bulge \citep{Zoccali08, Johnson13_offaxis, Bensby17}. 
However, we emphasize that the presence of vertical gradients does not necessarily rule out populations following a single star formation history. In ``upside-down'' formation scenarios, older populations form from gas with a larger scale height than younger populations, naturally producing vertical gradients even from a single enrichment track. Similarly, time-dependent dynamical heating could preferentially lift older, more-metal-poor populations to larger scale heights. Both scenarios would produce the observed vertical gradients without requiring multiple distinct star formation episodes at different heights \citep{Bird2013, Minchev2015}.
These gradients arise from spatially varying SFEs and gas accretion rates, potentially leading to an underestimation of the metal-poor tail in our modeled MDF by up to $\sim$10\%--20\% in normalized stellar mass fraction. 
Future multizone models, incorporating radial dependencies, could better reproduce these features \citep[e.g.,][]{Chiappini2001, Matteucci09}.
The absence of explicit dynamical modeling prevents the inclusion of bar-driven gas flows, which operate on $\sim$100 Myr timescales and funnel gas from the disk into the bulge, influencing both the MDF bimodality and [$\alpha$/Fe] knee position \citep{Athanassoula05, Wegg13, Ness16}. 
This manifests as a potential overestimate of the second-infall timescale by $\sim$0.5--1 Gyr, as bar-induced inflows could rapidly enrich the bulge without requiring extended star formation.
Hydrodynamical simulations coupled with GCE, such as those in \citet{Portail17} or \citet{Fragkoudi2020}, highlight the need for such integrations to disentangle secular from merger-driven evolution.

Although our adopted yield sets incorporate metallicity-dependent nucleosynthesis (for both massive stars and SNe~Ia), our one-zone, single-reservoir approach still enforces homogeneous mixing throughout the bulge. As a result, any spatial or temporal inhomogeneities in gas enrichment---including local variations in initial metallicity, yield output, or star formation history---are inevitably averaged out.  

Thus, while the yields evolve with progenitor metallicity, our model cannot capture metallicity gradients or localized enrichment episodes. This homogenization likely suppresses the intrinsic scatter and spatial dispersion in the abundance ratios, potentially smoothing out features such as a broadened [Fe/H] distribution or regional [$\alpha$/Fe] substructure.  

Future work involving multizone or chemodynamical models, which permit spatially resolved enrichment and mixing, would be required to explore how yield variations manifest in spatial gradients or localized abundance spreads.  

Furthermore, we do not explicitly model metallicity gradients as a function of galactic position or height above the plane, limiting our ability to reproduce the full range of spatial abundance variations observed in the bulge \citep{Queiroz2021, lucey25}. Nor do we account for stellar migration or radial churning, which can redistribute older, metal-poor stars from the inner disk into the bulge, potentially broadening the modeled AMR by 1--2 Gyr at fixed [Fe/H] \citep{Kubryk15, Minchev25}.

\section{Physical Analysis}
\label{sec:physical}

\begin{figure*}[ht!]
    \centering
    \includegraphics[width=\linewidth]{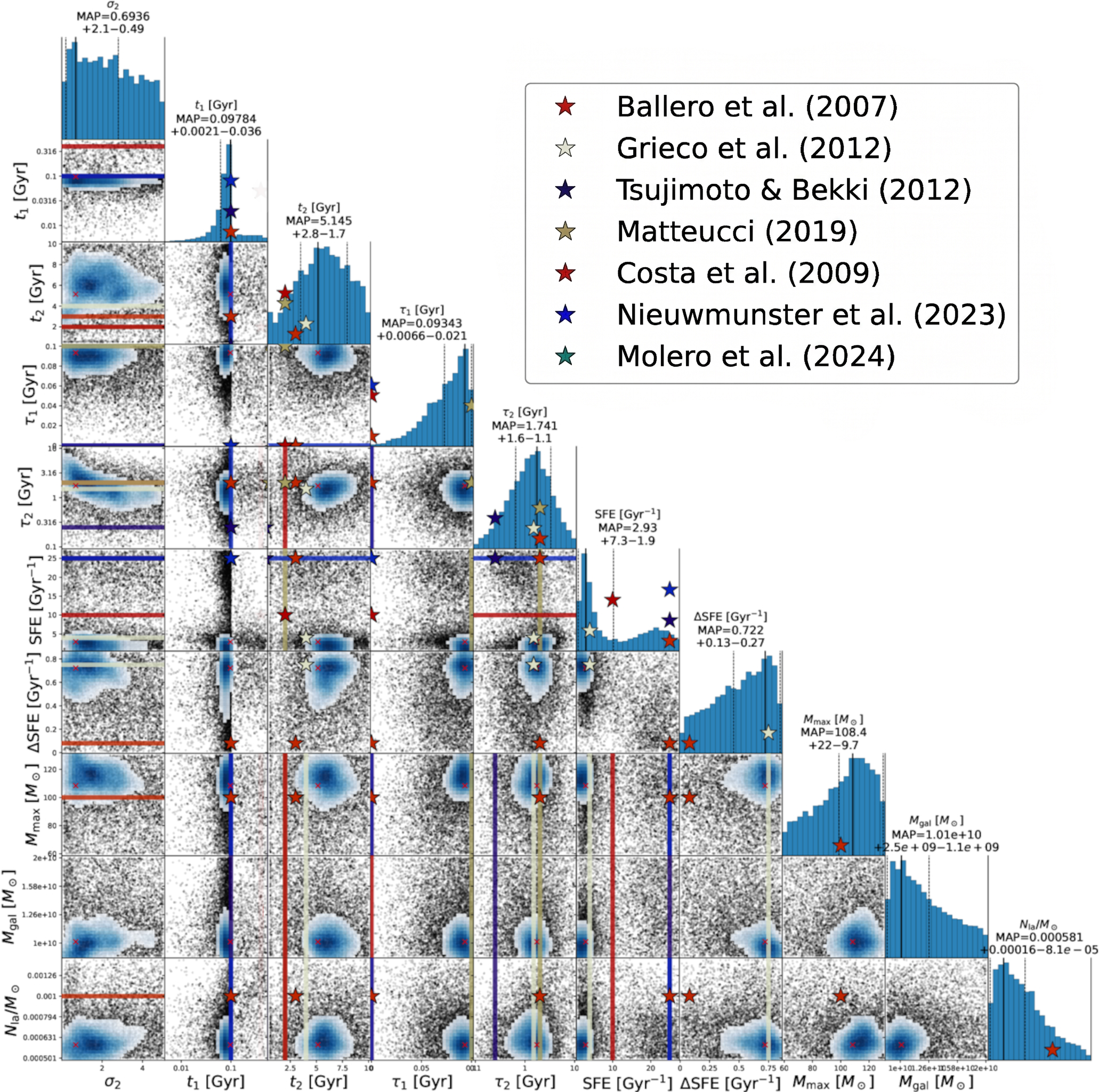}
    \caption{
    Posterior corner plot for the continuous model parameters.
    The diagonal panels show 1D marginalized distributions with MAP (solid line) and
    68\% HDI values (dashed lines).  
    The lower triangle panels show the MDF-weighted model ensemble as a background
    point cloud with the overlaid smoothed posterior density.
    The red crosses mark the MAP locations in each 2D plane.
    The colored stars and guide lines indicate the parameter choices adopted in
    previous bulge studies, as listed in Table~\ref{tab:lit_results}. 
    }
    \label{fig:big-corner}
\end{figure*}

The best-fit model for the bulge indicates an extremely rapid initial star formation episode.
We find that the first gas infall occurs at $t_1 \approx 0.1~\text{Gyr}$ after the birth of the Universe, with a very short infall duration of $\tau_1 \approx 0.09~\text{Gyr}$.
The bulge's primordial gas was accumulated and converted into stars on a timescale of order $10^8$~yr, which is a near-instantaneous collapse compared to the bulge's subsequent evolution. Such short collapse times naturally produce an early, $\alpha$-enhanced stellar population consistent with bulge age constraints.
The corresponding SFE is $\mathrm{SFE} \approx 2.93~\text{Gyr}^{-1}$ for this initial burst, which implies the gas would be largely depleted in only $\sim0.34$~Gyr ($1/\mathrm{SFE}$).
Such a high efficiency and brief timescale for the first episode are consistent with a classical rapid collapse scenario, producing a prompt enrichment of the bulge's oldest stellar population \citep{Matteucci90}.

By contrast, the second star formation episode is significantly delayed and more prolonged.
This follows directly from the posterior's preference for large $t_2$ and comparatively long $\tau_2$.
The model's second infall occurs at $t_2 \approx 5.1~\text{Gyr}$, several gigayears after the initial burst, and has an extended duration of $\tau_2 \approx 1.7~\text{Gyr}$.
This delayed infall can be interpreted as a secondary gas accretion (e.g., through a merger or cooling flow) around 8.6~Gyr ago, which reignites the star formation in the bulge.
The star formation in the second phase remains efficient but is moderately reduced relative to the initial burst. 
The best-fit $\Delta_{\rm SFE} \approx 0.72$ indicates that the second-phase SFE is about 28\% lower than in the first phase.
In absolute terms, the second episode's SFE is still on the order of $1.72~\text{Gyr}^{-1}$, sustaining rapid star formation, though not quite as extreme as the initial burst.

The inferred infall mass ratio between the two episodes is $\sigma_2 \approx 0.69$, implying that roughly 40\% of the bulge's stellar mass formed during the second, extended episode (with the remaining $\sim60\%$ produced in the initial collapse).
This substantial later contribution helps to explain the bulge's metal-rich population: the second infall dilutes the ISM and prolongs star formation, reproducing the observed metal-rich tail of the bulge MDF. Comparisons to results from previous bulge formation models are summarized in Table~\ref{tab:lit_results}.

\begin{table*}
    \centering

    \begin{tabular}{lcccccccccc}
        \hline
        Paper &
        $\sigma_2$ & Range &
        $t_1$ & Range &
        $t_2$ & Range &
        $\tau_1$ & Range &
        $\tau_2$ & Range \\
        (a)&
         &  &
        (Gyr) & &
        (Gyr) & &
        (Gyr) & &
        (Gyr) & \\        
        \hline
        [1] This study &
        0.694 & 0.01--5.0 &
        0.098 & 0.01--0.7 &
        5.14 & 0.1--10 &
        0.093 & 0.001--0.1 &
        1.74 & 0.1--10 \\

        [2] \citet{Grieco12} &
        -- & -- &
        0 & -- &
        2 & -- &
        0.10 & 0.10--0.30 &
        3 & -- \\

        [3] \citet{Molero2024} &
        0.4 & 0.1 -- 0.4 &
        0 & -- &
        Delay--0.25 & -- &
        0.10 & -- &
        $\sim 2$ & -- \\

        [4] \citet{Nieuwmunster23} &
        -- & -- &
        0 & -- &
        ``A delay'' & -- &
        0.40 & -- &
        2 & -- \\

        [5] \citet{Ballero07} &
        1 & -- &
        0 & -- &
        -- & -- &
        0.10 & 0.01--0.70 &
        2 & -- \\

        [6] \citet{Takuji2012} &
        0.5 & -- &
        0.3 & -- &
        1.5 & -- &
        1.0 & -- &
        4.0 & -- \\

        [7] \citet{Matteucci19} &
        0.5 & -- &
        0.5 & -- &
        0.75 & -- &
        0.1 & -- &
        0.1 & -- \\

        [8] \citet{Costa2009} &
        0.18 & -- &
        0.1 & -- &
        2.0 & -- &
        0.6 & -- &
        2.0 & -- \\        
        \hline
    \end{tabular}

    \begin{tabular}{lcccccccccc}
        \hline
        Paper &
        SFE$_1$ & Range &
        SFE$_2$ & Range &
        $M_{\max}$ & Range &
        $M_{\mathrm{Bulge}}$ & Range &
        $N_{\rm Ia}/M_\odot$ & Range \\
        (b) &
        (Gyr$^{-1}$) & &
        (Gyr$^{-1}$) & &
        ($M_\odot$) & &
        ($10^{10} M_\odot$) & &
        &  \\        
        \hline
        [1] &
        2.93 & 2--200 &
        2.12 & 0.01--17.0 &
        108.4 & 60--130 &
        1.01 & 0.9--2 &
        $5.81\times 10^{-4}$ & 0.0005--0.0015 \\

        [2] &
        25 & -- &
        2 & -- &
        100 & -- &
        2 & 2--2.28 &
        0.001 & -- \\

        [3] &
        25 &  -- &
        25 & -- &
        -- & -- &
        1 & -- &
        Iwamoto W7 & -- \\

        [4] &
        10 & -- &
        10 & -- &
        -- & -- &
        -- & -- &
        -- & -- \\

        [5] &
        20 & 2--200 &
        -- & -- &
        100 & -- &
        -- & -- &
        -- & -- \\

        [6] &
        4.0 & -- &
        3.0 & -- &
        -- & -- &
        -- & -- &
        -- & -- \\

        [7] &
        25.0 & -- &
        25.0 & -- &
        -- & -- &
        -- & -- &
        -- & -- \\

        [8] &
        -- & -- &
        -- & -- &
        -- & -- &
        -- & -- &
        -- & -- \\        
        \hline
    \end{tabular}


        





    
    \caption{Comparison of two-infall bulge models in this study and in the literature. 
    Table~\ref{tab:lit_results}(a) lists the infall mass ratio between the second and first episode ($\sigma_2$), onset times ($t_1$, $t_2$), and infall timescales ($\tau_1$, $\tau_2$). 
    Table~\ref{tab:lit_results}(b) lists the SFEs in the first and second episodes (SFE$_1$, SFE$_2$), the IMF upper mass ($M_{\max}$), bulge stellar mass ($M_{\mathrm{Bulge}}$), and SN\,Ia normalization ($N_{\rm Ia}/M_\odot$). 
    ``Range'' denotes the explored grid or adopted interval.}
\label{tab:lit_results}
\end{table*}

The literature values collected in Table \ref{tab:lit_results} reveal several systematic trends in two-infall bulge modeling that allow us to contextualize our results.
All previous studies favor a fast early bulge formation phase, followed by a later episode, but our solution combines a very rapid first infall with a comparatively late and moderately extended second infall.
Our posterior supports the standard two-phase bulge picture while shifting more of the late activity to later cosmic times than most prior models.

Our first infall ($t_1 \simeq 0.1$~Gyr, $\tau_1 \simeq 0.09$~Gyr) sits at the fast end of the literature range, and it is comparable to the short $\tau_1 \sim 0.1$~Gyr adopted by \citet{Grieco12}, \citet{Molero2024}, and \citet{Ballero07}.
The second episode in our model begins later ($t_2 \simeq 5.1$~Gyr) than in \citet{Grieco12} ($t_2 \sim 2$~Gyr), with an intermediate timescale ($\tau_2 \simeq 1.7$~Gyr, between the $\sim 1.5$--$3$~Gyr used by \citealt{Grieco12, Takuji2012, Nieuwmunster23}).
Our inferred mass ratio ($\sigma_2 \simeq 0.69$) implies that the second episode contributes $\sim 40\%$ of the bulge mass, intermediate between the smaller late contribution assumed by \citet{Molero2024} ($\sigma_2 \simeq 0.4$).
Our SFEs (SFE$_1 \simeq 2.9$~Gyr$^{-1}$ and SFE$_2 \simeq 2.1$~Gyr$^{-1}$) are high but well below the very intense bursts (SFE $\sim 20$--$25$~Gyr$^{-1}$) adopted by \citet{Grieco12}, \citet{Molero2024}, and \citet{Matteucci19}.
Our bulge stellar mass ($M_{\mathrm{Bulge}} \simeq 1.0\times 10^{10}\,M_\odot$) and SN\,Ia rate per unit mass ($N_{\rm Ia}/M_\odot \simeq 5.8\times 10^{-4}$) fall toward the lower end of the ranges assumed in previous studies. 
For example, \citet{Grieco12} adopt $M_{\mathrm{Bulge}} \sim 2\times 10^{10}\,M_\odot$ and $N_{\rm Ia}/M_\odot \sim 10^{-3}$, while our values are closer to the $M_{\mathrm{Bulge}} \sim 10^{10}\,M_\odot$ used by \citet{Molero2024}.

\subsection{The MDF}
\begin{figure*}[htb!]
\centering
\includegraphics[width=\linewidth]{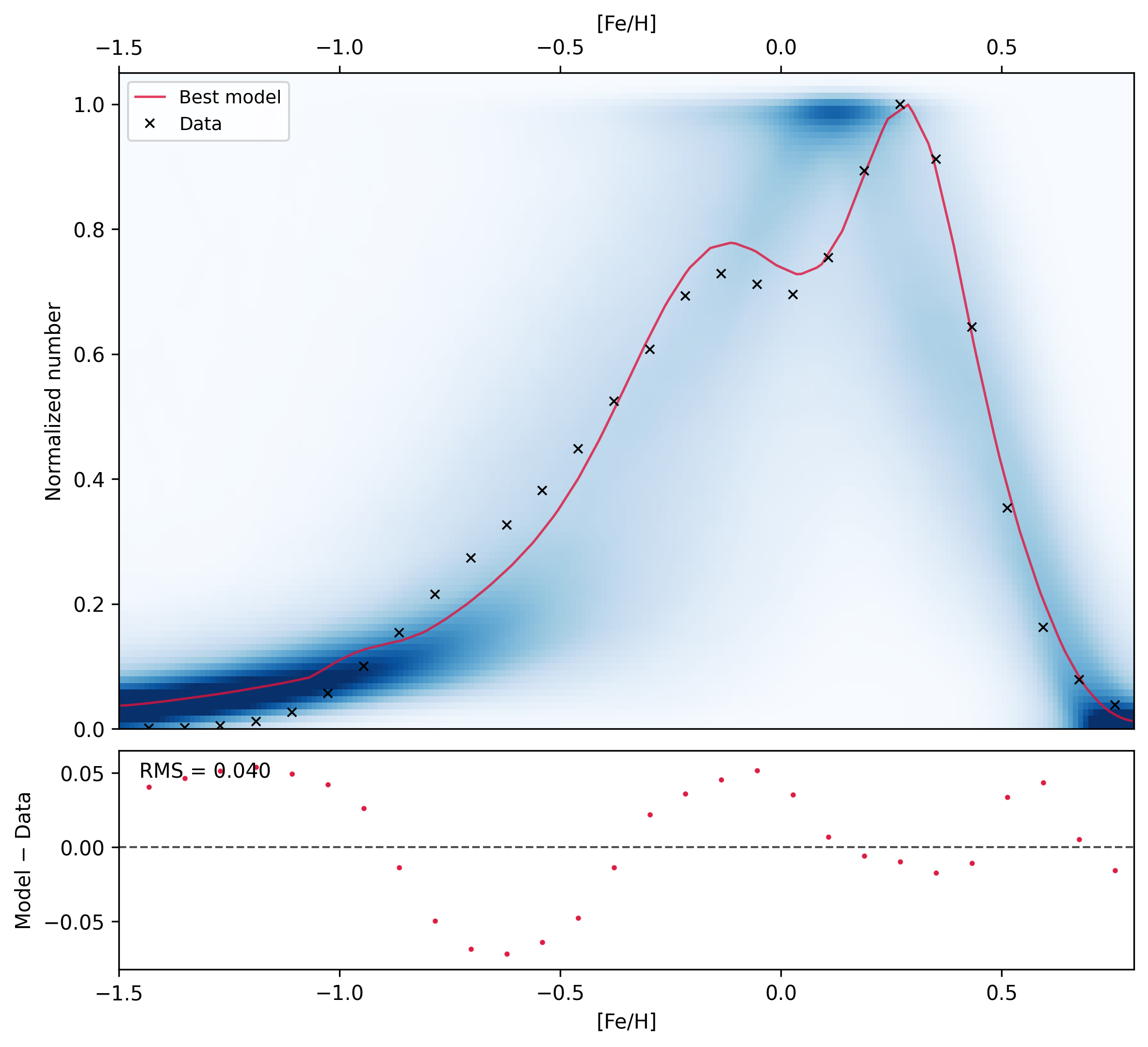}
\caption{MDF of the bulge for the best-fit two-infall model compared to the observed MDF. In the upper panel, the blue shading shows the posterior predictive distribution of model MDFs as a function of $\mathrm{[Fe/H]}$, the red curve marks the single best-fit model realization, and the black crosses show the empirically measured normalized MDF. The lower panel displays the residuals (model $-$ data) for the best-fit curve in each metallicity bin, with the overall rms residual of $\mathrm{rms}=0.040$.}
\label{fig:mdf-fit}
\end{figure*}

\vspace{0.2cm}

Figure \ref{fig:mdf-fit} shows how the posterior ensemble of models compares to the observed MDF.
The best-fit two-infall models successfully recover the observed MDF with high fidelity, achieving an rms residual of $0.040$.
The MDF encodes the integrated enrichment history of the bulge, so matching its shape is the primary aim of this study.
The models capture both the characteristic near-solar-metallicity peak and the extended metal-poor tail that defines the bulge population.
The two primary peaks---at [Fe/H] $\approx -0.3$ and +0.3---are reproduced well, supporting the two-infall method's ability to generate bimodal distributions through distinct enrichment episodes \citep{Bensby11,Hill11,Grieco12}.
The metal-poor peak reflects the rapid first infall, while the near-solar peak arises from the prolonged second episode after dilution and renewed star formation.
However, the models also tend to slightly overpredict the low-metallicity tail at [Fe/H] $\lesssim -1.0$. This is possibly due to observational selection effects that make metal-poor stars harder to detect in the crowded bulge fields, resulting in an underrepresented tail in the empirical data \citep{Johnson2022}.
Conversely, the metal-rich tail at [Fe/H] $> 0.3$ shows minor underprediction in some realizations, suggesting that the adopted infall prescription may slightly underestimate late-time gas accretion or that SN feedback effects require further refinement to match the observed spread.

\subsection{$\alpha$-elements}\label{sec:alpha_feh}
\begin{figure*}[htb!]
\centering
\includegraphics[width=\linewidth]{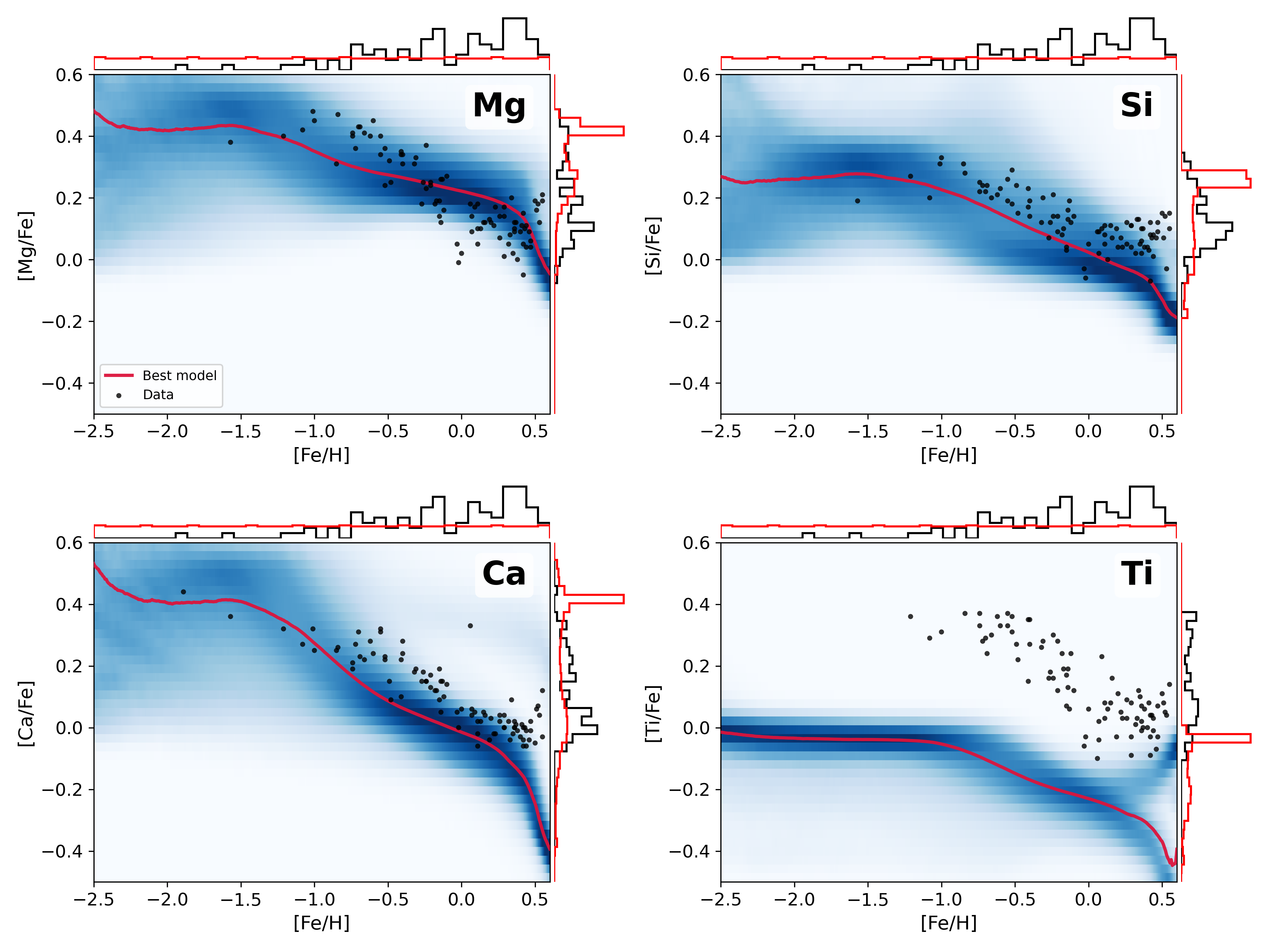}

\caption{Posterior [$\alpha$/Fe] [Fe/H] relations for the individual bulge $\alpha$-elements Mg, Si, Ca, and Ti for the best-fit two-infall model. In each panel, the blue shading shows the posterior predictive distribution of the model abundance ratios, the red curve traces the median (best-fit) model track, and the black points indicate the observed stellar abundances. The top and right insets give the corresponding 1D [Fe/H] and [$\alpha$/Fe] histograms for the data (black) and model (red).}
\label{fig:alpha-fit}
\end{figure*}

The $\alpha$-element abundances shown in Figure~\ref{fig:alpha-fit} provide constraints on the nucleosynthetic processes and timescales governing the bulge chemical evolution.
The best-fit two-infall models reasonably reproduce the observed [$\alpha$/Fe] sequences for Mg, Si, and Ca across the full metallicity range, capturing both the characteristic $\alpha$-enhancement at low [Fe/H] and the subsequent decline toward solar values at higher metallicities. This agreement is expected for elements primarily produced in massive stars.

For Mg, Si, and Ca, the models reproduce a reasonable canonical $\alpha$-element behavior expected from two-phase chemical evolution, where metal-poor stars ([Fe/H] $< -1.0$) exhibit $\alpha$-enhancements of $> 0.15$ dex, reflecting the dominance of core-collapse SNe during early rapid star formation \citep{Nomoto2013}.
The transition region around [Fe/H] $\sim -0.5$ shows the expected decline as SNe~Ia begin contributing substantial iron-peak elements \citep{Nomoto2013} and metal-rich populations ([Fe/H] $> 0.0$) converge toward solar abundance ratios within the range $-0.2 < [\alpha/\text{Fe}] < 0.2$, consistent with the delayed onset of iron production from the type Ia channel \citep{Matteucci09}.
Significant tension remains for Ti abundances, where the models systematically underpredict the observed enhancement across all metallicities.
This titanium deficit reflects a well-documented challenge in GCE modeling, with multiple studies showing that standard nucleosynthesis yields systematically underproduce Ti relative to observations \citep{Timmes1995, Prantzos2018, Kobayashi2020, Truemam2025}. 
This discrepancy is thought to arise from titanium being synthesized in a very thin $\alpha$-rich freeze-out layer during explosive Si burning, making the yield extremely sensitive to uncertain quantities such as the mass cut, electron fraction ($Y_e$), and multidimensional explosion dynamics.  
Because 1D stellar evolution and explosion models cannot accurately capture these narrow turbulence-dependent regions, they consistently underpredict Ti (\citealt{Woosley1995}; \citealt{Thielemann1996}; LC18).

Beyond individual element tracks, the overall $\alpha$-element distribution function (ADF)---the distribution of stars at different [$\alpha$/Fe] values---shows notable tension with observations. As is evident from Figure~\ref{fig:alpha-fit}, the best-fit model and much of the posterior ensemble underproduce low-[$\alpha$/Fe] populations relative to both the microlensed dwarf sample and the broader APOGEE ADF. This manifests as a deficit of stars with [$\alpha$/Fe] $\lesssim 0.1$ at [Fe/H] $> -0.3$. Several factors may contribute to this discrepancy: (i) the ADF was not used as a direct fitting constraint, only as validation; (ii) the one-zone approximation cannot capture the full diversity of enrichment histories that may produce the observed $\alpha$-element spread; (iii) selection-function differences between APOGEE and the microlensed sample complicate direct comparisons; and (iv) the model's single SFE prescription for each phase may oversimplify the true star formation history. Future work incorporating the ADF as a primary constraint---particularly the location of the [$\alpha$/Fe] knee and the relative populations of high- and low-$\alpha$ stars---would help diagnose and potentially resolve this tension.

\subsection{AMR}
\label{sec:amr}

\begin{figure*}[htb!]
\centering
\includegraphics[width=\linewidth]{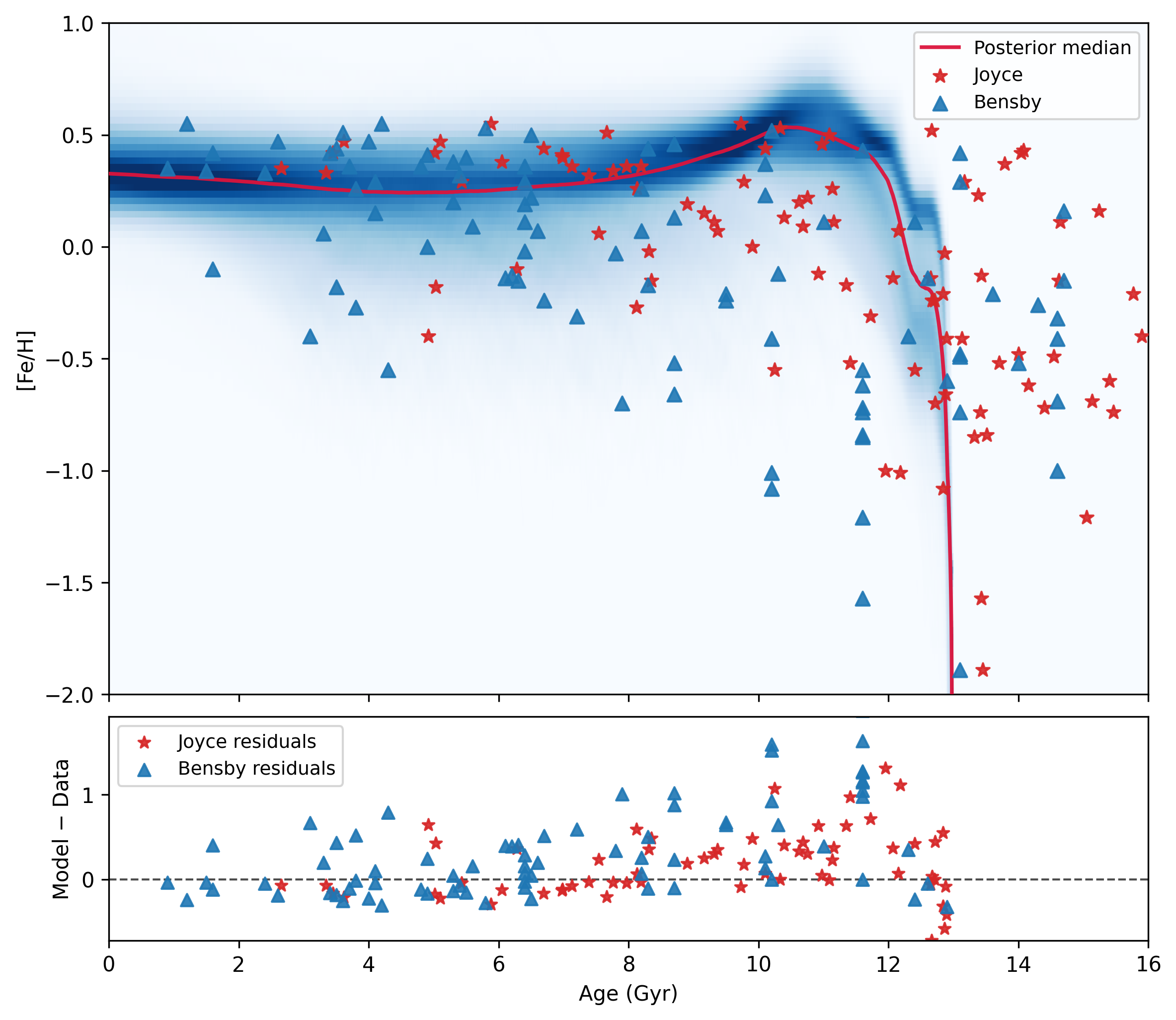}
\caption{
Posterior AMR for the bulge.
The blue density field shows the weighted posterior ensemble of chemically
acceptable models, while the solid red line traces the MAP model.
Individual stellar measurements are overplotted for comparison:
the \citet{Joyce2023} ages as red stars, and the \citet{Bensby17} ages as blue triangles.
The lower panel shows the residuals in [Fe/H] (model $-$ data) as a
function of age for both samples.
}
\label{fig:amr-fit}
\end{figure*}

The AMR allows us to probe the timing of enrichment and the relative contributions of the two infall episodes throughout the bulge formation history.
Figure~\ref{fig:amr-fit} compares the posterior AMR implied by our MDF-constrained models with the microlensed bulge dwarfs of \citet{Bensby17}, using the revised ages from \citet{Joyce2023} (red stars) and the original \citet{Bensby17} ages (blue triangles).
The posterior ensemble forms a relatively narrow high-metallicity ridge from $\sim 2$ to $\sim 11$~Gyr, with [Fe/H] remaining close to solar over most of this interval, then declining sharply at the oldest ages, reflecting a predominantly old bulge population.
This behavior is a direct consequence of the model's rapid early enrichment and sustained high metallicity over most of the bulge history.
The \citet{Joyce2023} ages for the \citet{Bensby17} stars lie closer to this ridge, with residuals typically within $\sim 0.1$ $0.2$~dex in [Fe/H].
In contrast, the original \citet{Bensby17} ages place many supersolar-metallicity stars younger than the model sequence.
For [Fe/H]~$>0$, many of the \citet{Bensby17} ages are displaced to younger values by $\Delta t \sim 2$--$3$~Gyr relative to the posterior ridge, which appears as a coherent trend in the residual panel at ages $\lesssim 6$--$7$~Gyr.

A physical interpretation that fits the AMR results is as follows.
The recovered parameters indicate an early, rapid first infall, with onset time $t_1 = 0.098$~Gyr and infall timescale $\tau_1 = 0.093$~Gyr (Table~\ref{tab:param_hdi_degeneracy}).
This first phase rapidly enriches the bulge ISM to near-solar [Fe/H] within the first $\sim 1$~Gyr, establishing the high-metallicity plateau seen at old ages in Figure~\ref{fig:amr-fit}.  
A second, delayed infall episode then replenishes the gas reservoir at intermediate ages.  
In the MAP solution, this occurs at $t_2 = 5.15$~Gyr, with a longer infall timescale $\tau_2 = 1.74$~Gyr and a mass ratio $\sigma_2 = 0.69$, so that the second infall contributes a substantial but subdominant fraction of the total accreted gas.  
This delayed gas supply slows any metallicity decline and maintains high [Fe/H] at intermediate ages ($t \sim 8$--$10$~Gyr), naturally accommodating the modest intermediate-age, metal-rich tail seen in the \citet{Joyce2023} ages without requiring a large population of very young stars.
The SFE in the first phase is high, with a MAP value of ${\rm SFE}_1 \simeq 2.93~{\rm Gyr}^{-1}$ and a 68\% HDI spanning $\sim 1$--$9~{\rm Gyr}^{-1}$, ensuring rapid early enrichment.  
The second episode proceeds with a reduced efficiency, characterized by $\delta_{\rm SFE} \simeq 0.722$ (HDI $\sim 0.39$--$0.85$), implying that the SFE during the later phase is $\sim 40\%$ lower than in the initial burst.  
This more gradual lower-efficiency star formation during the second infall allows SNe~Ia to contribute a larger fraction of the iron budget while stars are still forming, producing the lower [$\alpha$/Fe] ratios of the intermediate-age, metal-rich bulge population identified by \citet{Joyce2023}.
Taken together, this combination of short $\tau_1$, high early SFE, and mid-age gas replenishment with lower second-phase SFE is what simultaneously reproduces the MDF peak near solar metallicity, sustains the extended metal-rich tail at late times, and yields an AMR that is broadly consistent with the \citet{Joyce2023} ages and consistent with the presence of intermediate-age populations in the bar (e.g., the background bar--bulge Mira population in \citealt{Sanders2024}).

\begin{figure}[htb!]
\centering
\includegraphics[width=\linewidth]{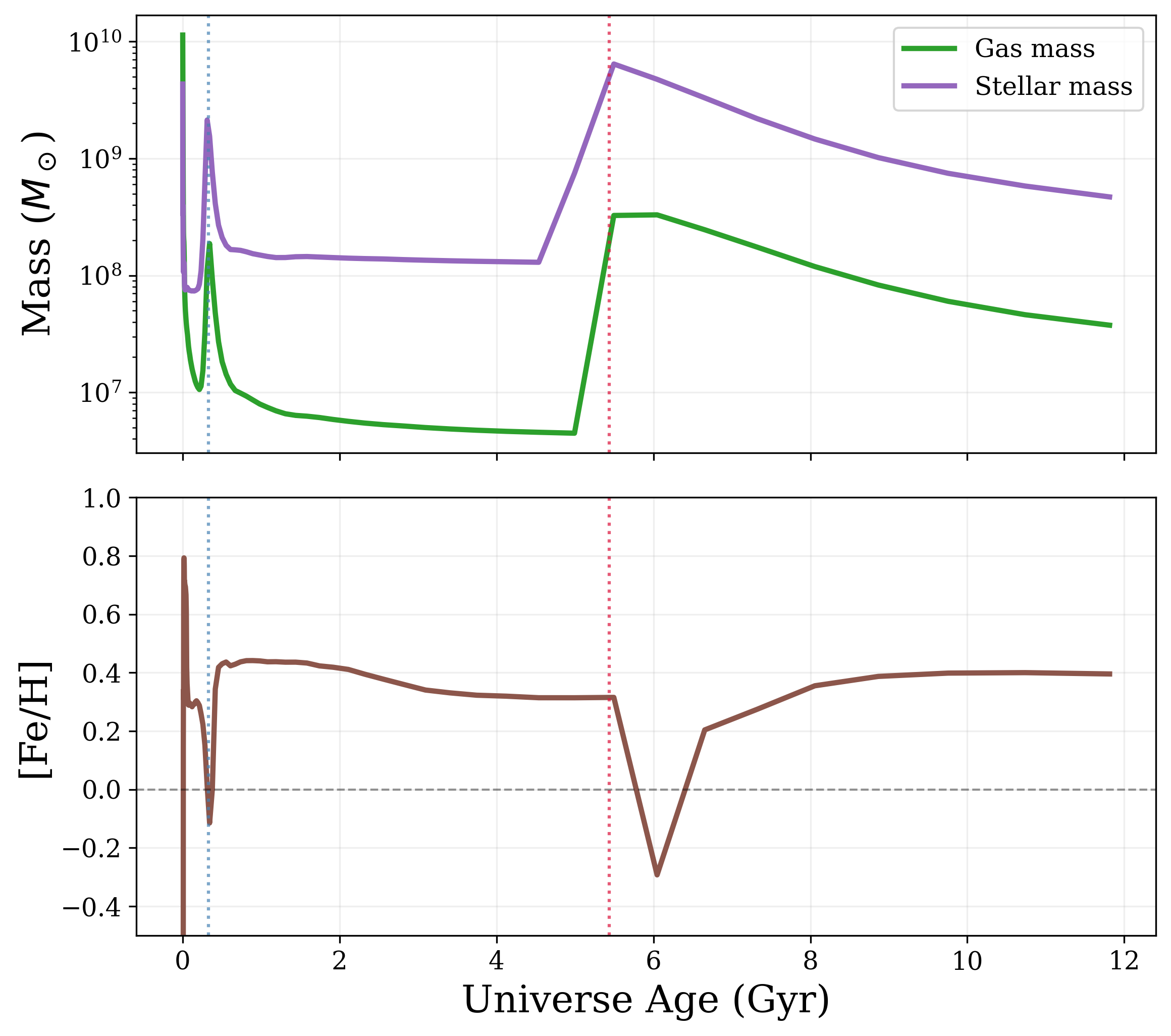}
\caption{
Gas reservoir mass (green) and stellar mass (purple) as functions of Universe age for the best-fit model (upper panel), and the corresponding ISM [Fe/H] evolution (lower panel). 
The vertical dotted lines mark the onset times of the first (blue) and second (red) infall episodes. 
The dip in ISM [Fe/H] at $\sim6$~Gyr corresponds to the dilution of the gas reservoir by the second infall before renewed star formation can reenrich the ISM, providing a physical explanation for the tension seen in the AMR at $\sim9$--$12$~Gyr lookback time. The dashed horizontal line marks solar metallicity.
}
\label{fig:gas-metal}
\end{figure}

We acknowledge that the model AMR exhibits a metallicity dip of $\sim 0.2$~dex in the $\sim 9$--$12$~Gyr range, corresponding to the transition between infall episodes. This feature arises from gas dilution as the second infall introduces lower-metallicity material before renewed star formation can reenrich the ISM. Notably, this dip is not observed in any dataset with reliable ages, including the microlensed observations. This discrepancy likely reflects: (i) the simplified prescription for accreting gas metallicity in our model, which draws from the STELLAB library rather than tracking a self-consistent CGM enrichment history; (ii) the abrupt transition between infall phases, whereas physical gas accretion may be more gradual; and (iii) the one-zone assumption, which cannot capture spatial inhomogeneities that might buffer metallicity changes. Future models that track the metallicity evolution of accreting gas self-consistently, or that allow for continuous rather than two-phase infall prescriptions, may help resolve this tension. Figure~\ref{fig:gas-metal} shows the gas reservoir mass and ISM [Fe/H] as functions of time for the best-fit model, providing direct diagnostic context for this dip.

The posterior AMR is less compatible with the ages reported in the original \citet{Bensby17} analysis for supersolar-metallicity stars, many of which appear younger than predicted by our two-infall histories, despite their measured [Fe/H] being consistent with the model.

\begin{figure*}[htb!]
\centering
\includegraphics[width=\linewidth]{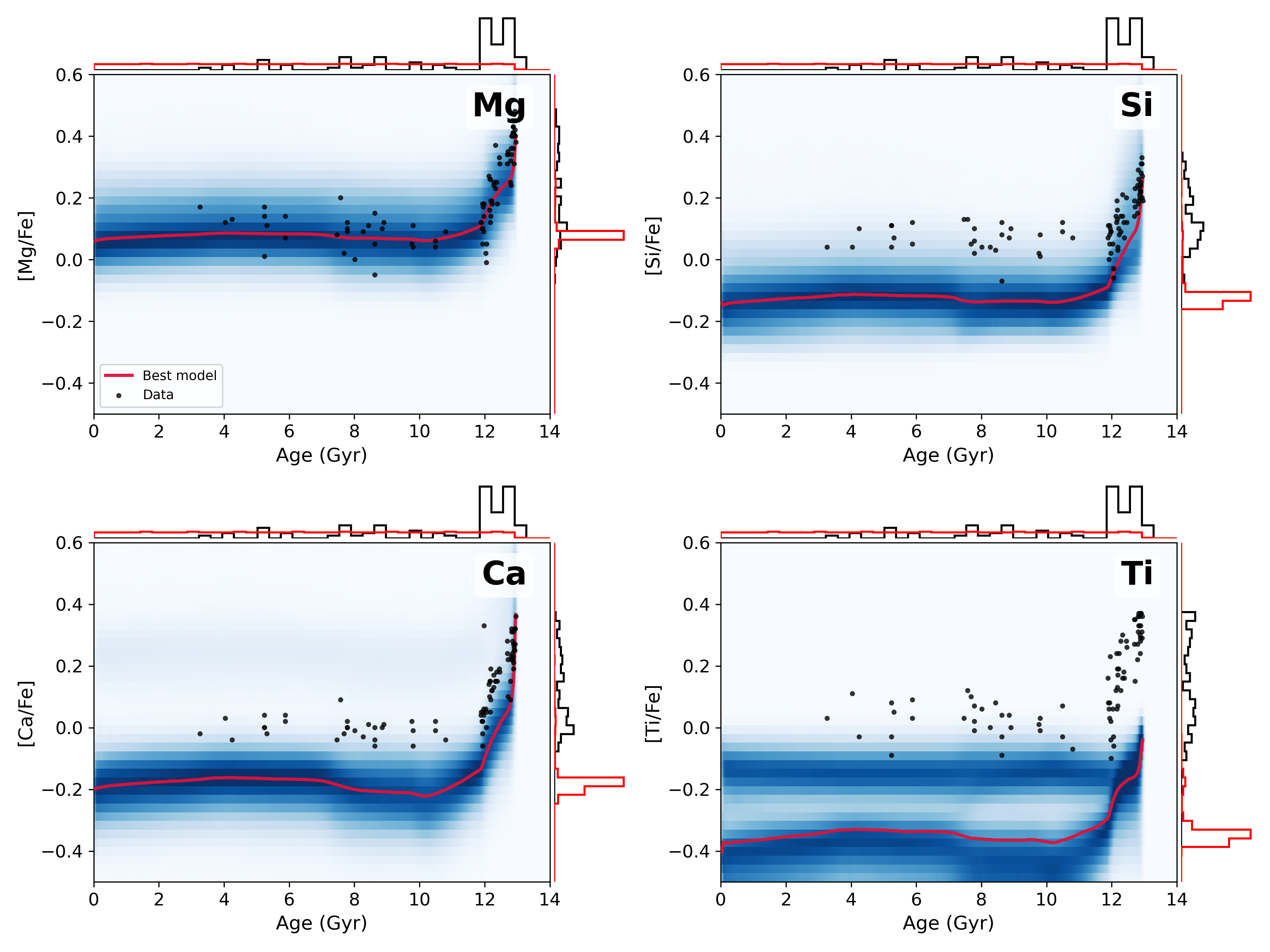}
\caption{
Posterior age--abundance relations obtained by mapping each model's [$\alpha$/Fe]--[Fe/H] track through the median posterior AMR.
The blue shading indicates the weighted posterior density, and the solid red curve shows the best model.  
The observed bulge abundances are transformed into age using the same AMR mapping and shown as the black points.  
The panels show the intermediate-age metal-rich population suggested by the AMR occupies the low-[$\alpha$/Fe] sequence at ages $\sim 6$--$10$~Gyr, while the oldest stars remain on the high-[$\alpha$/Fe] plateau.  
}
\label{fig:amr-alpha}
\end{figure*}

Figure~\ref{fig:amr-alpha} shows the [$\alpha$/Fe]--age relations generated by mapping each model's [$\alpha$/Fe]--[Fe/H] track through the posterior AMR.  
All four panels show a long, nearly flat, low-[$\alpha$/Fe] sequence from ages $\sim 2$--$10$~Gyr, which is traced by both the posterior and the transformed data.  
At the oldest ages ($\gtrsim 13$~Gyr), the posterior exhibits a sharp rise in [$\alpha$/Fe], reflecting the chemically old high-$\alpha$ population.  
The data follow the same overall structure, with the majority of stars occupying the low-[$\alpha$/Fe] band at intermediate ages and a smaller subset extending toward higher [$\alpha$/Fe] at the oldest ages.

\subsection{Interpretation}

In the following subsections, we consider the physical implications of our models' preferences for certain parameter combinations

\subsubsection{First-infall Timing}
The best-fit $t_1$ is very short (posterior peak $\sim$0.10\,Gyr), indicating that the initial collapse of gas into the bulge occurred extremely early in the Galaxy's history.
Such a short formation timescale means the early bulge underwent a prompt starburst, consistent with classical models that require a fast bulge formation to match its chemistry \citep{Ballero07, Grieco12}.
Correspondingly, the inferred SFE in this first phase is extremely high (best fit $\simeq 2.9$~Gyr$^{-1}$; see Figure~\ref{fig:big-corner}), implying gas was converted into stars on a timescale of only $\sim 0.34$~Gyr.
This high SFE and rapid collapse are in line with previous studies, which found that reproducing the bulge's [Fe/H] distribution and enhanced [$\alpha$/Fe] ratios requires a burst of star formation that is much more intense and short-lived than in the solar neighborhood.
Chemical evolution models have long predicted that the bulge formed in a brief, vigorous, star-forming burst (timescale $\lesssim 0.5$~Gyr) with an efficiency $\sim$10--20 times that of the disk \citep{Grieco12, Athanassoula05, Nieuwmunster23}.

\subsubsection{Second-infall Timing and the Origin of the Late Gas Supply}

The posterior allows us to investigate which physical mechanism plausibly supplied the late gas that fuels the second infall. The second infall episode is inferred to have occurred at $t_2 \simeq 5.1$~Gyr after the onset of the bulge evolution, with a broad 68\% HDI spanning $t_2 \simeq 3.2$--$8.4$~Gyr (Table~\ref{tab:param_hdi_degeneracy}). Assuming a present-day cosmic age of $13.8$~Gyr, this corresponds to a lookback time of $\sim 8.7$~Gyr for the MAP solution, with an allowed range of roughly $5.4$--$10.6$~Gyr. Thus, the second episode is clearly separated from the near-instantaneous first collapse, but its onset is only weakly constrained in absolute cosmic time.

A plausible explanation for this secondary infall could be found in the GSE event. Chemodynamical studies typically place the last major merger between $\sim 8$ and $11$~Gyr ago, with most GSE stars forming $\sim 10$--$11$~Gyr ago and the merger depositing $\sim 5\times10^{10}\,M_\odot$ of material into the Milky Way halo and inner regions \citep[e.g.,][]{Belokurov2018,Chaplin2020,Feuillet2021,Xiang2022}. Our $t_2$ posterior overlaps the younger end of this range.
The primary basis for associating this second infall with a GSE-like event is this timing coincidence: our inferred onset time ($t_2 \sim 5.1$~Gyr) aligns with the merger epoch derived from independent chemodynamical studies.
If GSE---or a similar early massive accretion event---drove a substantial inflow of gas toward the inner Galaxy, the second infall in our model may be interpreted as the chemically processed tail of that externally supplied gas reaching the bulge on Gyr timescales \citep{Ryde2025}. In that picture, the delayed, moderately extended influx ($\tau_2 \sim 1.7$~Gyr) reflects how long it takes for merger-induced torques and subsequent disk settling to deliver gas to the central kiloparsec region of the Milky Way.

Secular bar-driven inflows provide an equally plausible channel. Once a stellar bar forms, gravitational torques efficiently funnel disk gas inward along $x_1/x_2$ orbits on timescales of a few $10^8$~yr \citep[e.g.,][]{Fragkoudi2020}. Dynamical models broadly allow the Milky Way bar to form several Gyr after the earliest star formation, with typical estimates placing the bar formation sometime in the last $\sim 6$--$9$~Gyr, depending on the assumed disk mass and feedback history. Our $t_2$ range is fully consistent with a scenario in which the second infall is primarily fueled by bar-driven inflows of preenriched inner-disk gas (originating from the thick and/or early thin disk), rather than by lower-metallicity material directly associated with GSE, both of which were already in place by the epoch at which bar-driven inflows become significant.

Given the breadth of the $t_2$, $\tau_2$ posterior, hybrid scenarios are also physically plausible. A third possibility is that the second infall in our one-zone model is an effective parameterization of a composite late gas supply. Early external accretion (e.g., GSE) sets up a massive turbulent inner disk, and the subsequent bar then channels a fraction of that gas, together with chemically evolved thin-disk gas (already forming by $\sim$8--10 Gyr ago), into the bulge over several gigayears.

Within the broad $t_2$ and $\tau_2$ posteriors, our model cannot uniquely distinguish between an explicitly merger-driven second episode and a primarily secular bar-mediated inflow. Despite this, it is evident that some substantial late gas replenishment is required. Models with strictly negligible second infall fail to reproduce the strong solar/supersolar MDF peak and the high-[Fe/H] tail, whereas a delayed, moderately extended second infall of preenriched gas naturally builds the younger metal-rich component while remaining consistent with the $\alpha$-element and age--metallicity constraints.

\subsubsection{Implications of a Second Infall}

The parameter that controls its relative mass contribution is the infall mass ratio, $\sigma_2$, defined as the ratio of the second to the first-infall mass.
So, $\sigma_2$ governs the fraction of metal-rich stars and the height of the solar-metallicity peak. For a total accreted gas mass $M_{\text{tot}}$, the individual episode masses are
\begin{equation}
    M_1 = \frac{M_{\text{tot}}}{1 + \sigma_2}, \qquad
    M_2 = \sigma_2\, M_1 ,
\end{equation}
so that $\sigma_2 = 1$ corresponds to equal-mass episodes, $\sigma_2 > 1$ to a dominant second infall, and $\sigma_2 < 1$ to a dominant first infall.  
In the HPD region, the distribution of $\sigma_2$ peaks at the MAP value $\sigma_2 \simeq 0.69$, with typical values in the range $\sim 0.57$--$0.89$, implying that the second episode carries a substantial but not uniquely dominant fraction of the total accreted gas.

At the MAP, the first infall supplies $\sim 59\%$ of the gas and the second $\sim 41\%$; within the quoted range, the second episode contributes roughly $\sim 36\%$--$47\%$ of the total mass. This further supports a composite bulge formation scenario in which an early rapid collapse builds most of the stellar mass, but a later gas supply is still required to match the MDF and AMR.  
Because the MAP solution has the first episode dominating the mass budget, it remains compatible with a classical fast-formation bulge. Importantly, ``classical'' here refers to the rapid formation timescale, rather than a strictly metal-poor population. The nonzero second infall is best interpreted as adding a younger, more-metal-rich component on top of an already largely assembled old bulge, rather than rebuilding the bulge from scratch.  
The observed dominance of old stars (with a predominantly old bulge population, with the bulk of the microlensed sample older than $\sim 10$ Gyr; \citealt{Joyce2023}) is consistent with this picture, in which the first episode provides the majority of the mass, and even at the high end of the credible $\sigma_2$ range, the second infall remains a substantial but subdominant contributor. Its more extended timescale and lower SFE allow it to generate a younger, metal-rich population without obviously violating age constraints.

The joint posterior further shows that $\sigma_2$ is entangled with the timing and efficiency of the second phase. In Figure~\ref{fig:big-corner}, $\sigma_2$ forms broad, elongated ridges with ${\rm SFE}$, $\Delta{\rm SFE}$, and $t_2$, indicating that multiple combinations of second-episode mass fraction, onset time, duration, and SFE can reproduce the MDF and [$\alpha$/Fe] trends. Increasing $\sigma_2$ (more late gas) can be compensated for by a lower $\Delta{\rm SFE}$ (less efficient star formation), by shifting $t_2$, so that enrichment has more or less time to proceed, or by adjusting $\tau_2$, to spread the enrichment over a longer or shorter interval, and vice versa. As a result, the data constrain the existence and rough scale of a second infall, but do not uniquely fix its exact mass fraction or timing. Across the posterior, models with an effectively negligible second infall ($\sigma_2 \rightarrow 0$) fail to reproduce the strong solar and supersolar MDF peak and the high-[Fe/H] tail, whereas both moderate ($\sigma_2 \lesssim 1$) and somewhat larger ($\sigma_2 \gtrsim 1$) second-infall fractions can build the metal-rich peak, provided their SFE and timing are adjusted appropriately. The precise value of $\sigma_2$ is less tightly pinned down than the requirement that a nonzero second infall exists.

\subsubsection{Star Formation in the Second Episode ($\Delta_{\rm SFE}$ and $\sigma_2$)}

Given this substantial but subdominant late gas supply, the character of the star formation in the second episode is set by the SFE contrast parameter, $\Delta_{\rm SFE}$ (or $\delta_{\rm SFE}$), which scales the efficiency in the second phase relative to the first.
The posterior favors $\delta_{\rm SFE} \approx 0.72$, implying that the SFE during the second infall is roughly 28\% lower than in the initial burst.  
For typical first-episode efficiencies in the range ${\rm SFE}_1 \sim 2.9$--$4.2~{\rm Gyr}^{-1}$, this corresponds to ${\rm SFE}_2 \sim 2.1$--$3.0~{\rm Gyr}^{-1}$, i.e., still a vigorous but noticeably milder mode of star formation. 
Combined with the nonzero second-infall duration ($\tau_2 \sim 1.7$~Gyr), this reduced efficiency means that the second generation of bulge stars forms less intensely per unit gas mass and over a more extended period than the first, even though the total late-time gas mass (set by $\sigma_2$) remains significant.

This change in efficiency has direct chemical consequences. A more gradual, lower-SFE star formation during the second infall allows a larger fraction of SNe~Ia to explode while stars are still being formed, increasing the iron contribution from the delayed channel relative to the $\alpha$-elements from core-collapse events.
In the model, this naturally drives lower [$\alpha$/Fe] ratios in the younger, metal-rich bulge stars, matching the observed downturn in [$\alpha$/Fe] at high [Fe/H] and the location of the ``knee'' in [$\alpha$/Fe]--[Fe/H] near solar metallicity.  
The posterior thus favors a configuration in which an initial, very high-efficiency burst rapidly builds most of the bulge mass and establishes the $\alpha$-enhanced, metal-poor population, while a later, somewhat less efficient but still rapid second phase---fed by the nonzero $\sigma_2$---extends the star formation, boosts the metal-rich tail of the MDF, and imprints the low-[$\alpha$/Fe] sequence.  
This combination of a violent first burst followed by a milder, extended second episode is fully consistent with the abundance pattern constraints discussed above and with independent bulge chemical evolution studies.

\section{Conclusions}
A two-infall chemical evolution model suggests that the first infall episode occurs rapidly ($t_1 \approx 0.1$ Gyr, $\tau_1 \approx 0.09$ Gyr), forming $\sim60$\% of the bulge mass with high SFE ($\sim 2.9$ Gyr$^{-1}$), consistent with an early, rapid collapse, while the second infall episode is delayed ($t_2 \approx 5.1$ Gyr, $\tau_2 \approx 1.7$ Gyr) and moderately extended, contributing $\sim40$\% of the stellar mass with reduced SFE ($\sim 2.1$ Gyr$^{-1}$).

Using N'OMEGA+ with a GA to explore a 15D parameter space, our model reproduces the bulge's MDF, AMR, and $\alpha$-element patterns with high fidelity.
We find that the second infall is chemically required to produce the supersolar MDF peak and the low-[$\alpha$/Fe] population; models without a nonzero second episode cannot fit the data, even in the presence of degeneracies among $\sigma_2$, $t_2$, $\tau_2$, and $\Delta$SFE. The lower SFE and longer $\tau_2$ during this phase allow extended iron enrichment from SNe~Ia, generating the observed [$\alpha$/Fe] downturn. 
In this prescription, the first episode explains the $\alpha$-enhanced, metal-poor population, while the delayed second episode builds the younger, metal-rich tail and simultaneously reproduces the MDF bimodality, the [$\alpha$/Fe] knee position, and the \citet{Joyce2023} AMR without imposing direct age constraints.

Residual tensions remain with the supersolar ages and Ti abundances reported by \citet{Bensby17}, highlighting the need for improved stellar models and age determinations. 
A direct follow-up to this work will incorporate the AMR as a fitting constraint alongside the MDF, which we expect will better constrain the timing and character of the second infall episode and reduce the tension with the observed age--metallicity distribution.

PCA and MI analyses show that several parameter combinations can preserve the MDF and [$\alpha$/Fe] trends, and that categorical choices (IMF, yield set, and SN~Ia implementation) are not uniquely determined but are largely absorbed through compensating shifts in the continuous parameters. 
Even so, the requirement for two distinct infall episodes is robust, and the recovered SFE values are high but sub-bursting---lower than the 20--25~Gyr$^{-1}$ assumed in some previous bulge models and more consistent with modern observational constraints.

The bulge likely formed in a hybrid scenario: early rapid collapse followed by moderate secular or merger-driven gas inflow, without requiring a third distinct episode.
Limitations include the one-zone assumption, fixed yields, lack of dynamics, and no explicit modeling of radial migration or spatial gradients.
However, despite the simplifications, the two-infall + reduced-SFE model reproduces all the key observables and provides a coherent, physically motivated picture of bulge formation.

This study demonstrates the Milky Way bulge as a critical laboratory for disentangling the relationship between rapid high-redshift collapse and prolonged secular evolution in spiral galaxies. 
Our results demonstrate that reproducing the chemical profile of the bulge is attainable through a hybrid formation history. 
These chemical constraints provide a benchmark for future work, which must move beyond one-zone approximations to spatially resolved chemodynamical simulations capable of capturing radial and vertical gradients.

\section{Future Work}

A number of forthcoming observational advances and modeling improvements could significantly strengthen the constraints on bulge formation scenarios and help discriminate between competing hypotheses.

First, new and upcoming surveys such as the Nancy Grace Roman Space Telescope Galactic Bulge Time-Domain Survey (GBTDS) will provide deep, high-cadence, near-IR photometry---including grism spectroscopy ``snapshots''---for hundreds of millions of stars in the bulge region \citep{Gaudi2022}.  
These data will enable asteroseismic age estimates for large samples of red giant stars \citep{Huber2023, Weiss2025} and more robust metallicity and kinematic measurements, thereby providing far more precise MDFs and abundance--age distributions than are currently possible. 
Combined with long-time baseline surveys such as the Vera C. Rubin Observatory Legacy Survey of Space and Time (LSST), the joint optical and near-IR coverage will more effectively pierce dust extinction (with LSST reaching single-visit depths of $\sim24$--$25$~mag in the optical bands and up to $\sim27$~mag in the 10~yr coadds) and crowding toward the Galactic center (with a pixel scale of $0.2$ pixel$^{-1}$), increasing the completeness and stellar sampling depth \citep{Gonzalez2018, Street2018}.

A spatially resolved approach to modeling the bulge---splitting the bulge volume into multiple zones (e.g., by galactocentric radius and/or latitudinal binning)---would allow the reconstruction of radial and vertical metallicity gradients, localized enrichment episodes, and possible structural subpopulations.  
Observational metallicity maps like those derived from the VVV Survey and Two Micron All Sky Survey (2MASS) photometry already show clear vertical metallicity gradients across the bulge \citep{Gonzalez2013, Ness2016, Johnson2022}.  

Also, integrating chemical evolution modeling with dynamical or hydrodynamical simulations---allowing gas flows, radial mixing, bar-induced inflows, and noninstantaneous mixing---will yield more realistic predictions for the bulge's chemical and structural evolution.  Such models will be essential for predicting and comparing spatial abundance gradients, $\alpha$-element substructures, and AMRs as a function of position in the bulge.  Furthermore, combined chemical and hydrodynamical modeling will be important for deciphering the role bar formation timing plays in chemical enrichment.  The results presented here indicate a majority mass fraction forming at very early times, likely before the bar.  However, the slowly rotating, high-velocity component of the bulge is found to be relatively small \citep{Kunder16, Arentsen20}, while a majority of bulge stars exhibit bar-like kinematics \citep[e.g.,][]{Kunder12, Marchetti24}.  We expect that a combined approach will yield deeper insight into reconciling the interplay between chemistry and kinematics.

Finally, future iterations of this analysis could significantly reduce computational costs by adopting sparse-grid or emulator-based approaches from the design-of-computer-experiments literature \citep{Fang2005, Santner2013}, including Gaussian Process emulators that have been applied to high-dimensional astrophysical inference problems \citep{Kaufman2011, Wibking2020}.

A dedicated follow-up incorporating the AMR as a direct fitting constraint alongside 
the MDF is a natural next step, but rather than rerunning the present suite of models 
against the currently available, sparsely sampled AMR, we are instead awaiting the 
substantially more precise and spatially resolved AMR that Roman's 
GBTDS will deliver. M. Joyce et al.\ (2026, in preparation) will demonstrate a method by which the age posteriors for hundreds of thousands to millions of bulge stars can be computed using this dataset, providing a means of replacing the present-day MDF as the principal observable against which GCE models are compared. Using a 
Roman-informed AMR as a simultaneous target observable alongside the MDF will break 
the parameter degeneracies identified in this work---particularly the covariances between 
$t_2$, $\tau_2$, $\sigma_2$, and SFE that control the timing and character of the 
second infall episode---and will constitute the primary motivation for the next generation 
of bulge GCE modeling.

\begin{acknowledgments}
The authors wish to thank the University of Wyoming Advanced Research Computing Center (ARCC).
N.M. wishes to thank Katherine Miller for Figure~\ref{fig:intro-cartoon}.
M.J. wishes to thank John Bourke for typesetting.
This manuscript made use of the NASA ADS library.
\end{acknowledgments}

\software{OMEGA+/N'OMEGA+, Python, numpy, DEMC, pandas, BASH, LaTeX, ChatGPT, Gemini.}


\appendix
    
\section{Validation of \GAs{} with DEMC against Classical MCMC}

We ran an MCMC for each unique set of categorical parameters (outlined in Table \ref{tab:parameter_space}) and combined their posteriors.

\begin{figure}
    \centering
    \includegraphics[width=0.75\linewidth]{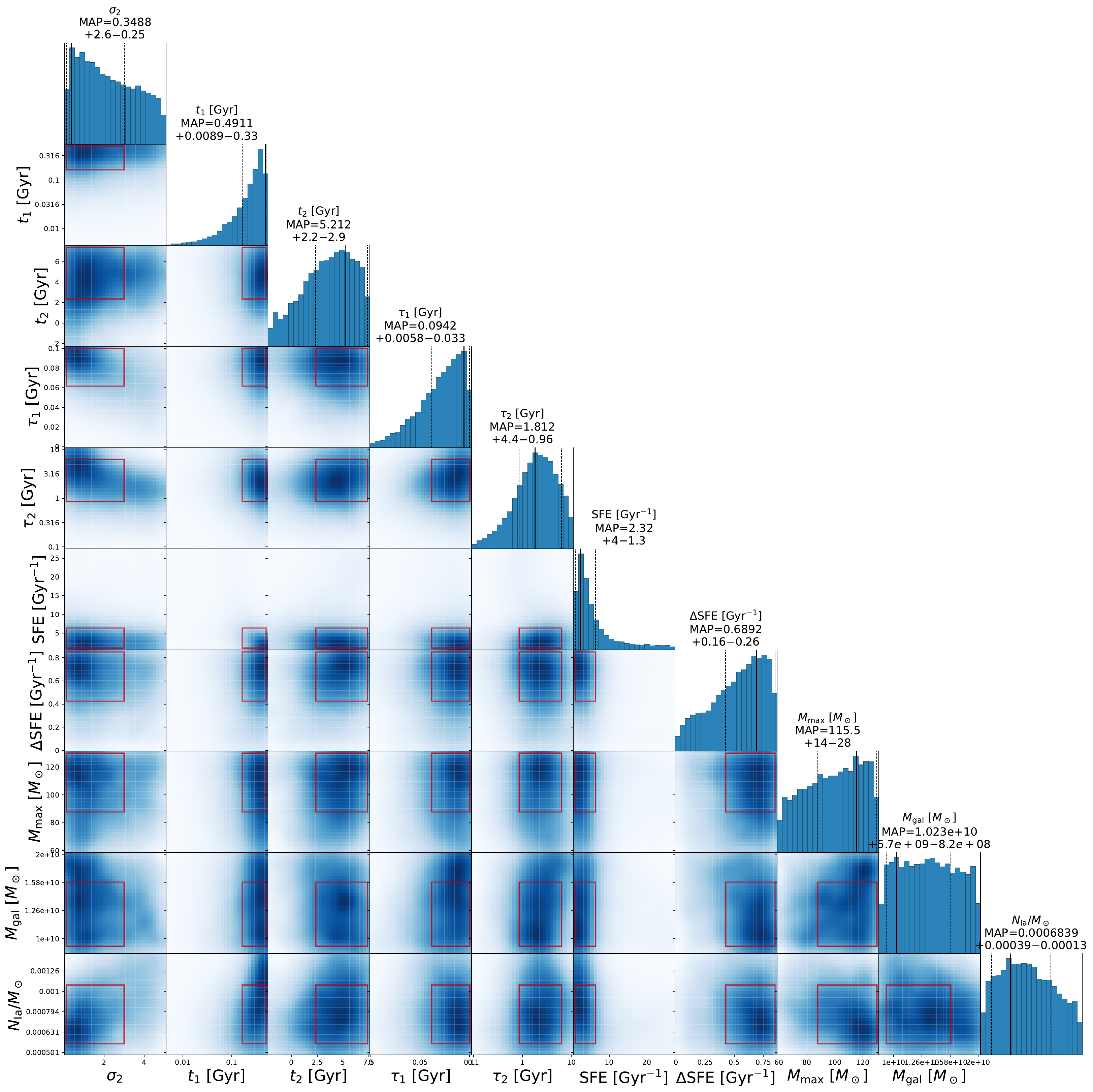}
\caption{
Corner plot of the joint posterior distribution in the continuous two-infall parameters obtained by combining 288 independent MCMC runs, one for each unique choice of categorical model ingredients (Table~\ref{tab:parameter_space}).
The blue histograms along the diagonal show the 1D marginalized
posteriors, with the MAP value and 68\% HDI annotated for each parameter. 
The off-diagonal panels display the corresponding 2D marginalized
distributions, with the darker shading indicating higher probability density and the red rectangles marking the MAP point and its projected 68\% HDI in each
parameter pair.
}

    \label{fig:placeholder}
\end{figure}

To check whether the GA with DEMC moves is sampling the same posterior as a conventional MCMC, we repeated the analysis with independent MCMC runs and compared the resulting posteriors.

For every distinct choice of categorical parameters, we launched a separate MCMC run in the continuous parameters.  
Each run used the same likelihood, priors, and parameter ranges as in the GA+DEMC analysis.
After discarding an initial burn-in segment, the chains were concatenated over all categorical combinations.  
The resulting catalog of MCMC samples was weighted in exactly the same way as the GA+DEMC models (using the MDF-based loss) and converted into a posterior.

From this combined MCMC posterior, we computed the same summary statistics as in the main text: MAP estimates and 68\% HDI limits for each parameter, and pairwise degeneracy metrics.

The 1D MCMC marginals agree with the GA+DEMC results to within the sampling noise, and the same dominant degeneracies appear in the 2D projections.  
Visually, the MCMC corner plot in Figure~\ref{fig:placeholder} is similar to the GA+DEMC posterior shown in Figure~\ref{fig:param_correlations}, and the MAP/HDI numbers in Table~\ref{tab:param_hdi_degeneracy} are reproduced by the MCMC chains.  
Table~\ref{tab:ga_mcmc_cost} summarizes the computational costs of the GA+DEMC and MCMC approaches.
This demonstrates that the GA with DEMC moves is tracing the same posterior distribution as a standard MCMC.

\begin{table}
    \centering
        \begin{tabular}{lcccc}
        \hline
        Method & No. of Model Evaluations & Wall Time (hr) & CPU Hours & Notes \\
        \hline
        MCMC    & 9,437,184 & 5016 & 642,048 & 288 chains combined \\
        GA+DEMC & 262,144 & 1056 & 135,168 & 16 GA runs combined \\
        GA+DEMC & 16,384 & 66 & 8448 & 1 GA run \\
        \hline
    \end{tabular}
    \caption{Comparison of Computational Costs for the GA+DEMC Analysis and the Pure MCMC Runs.}
    \label{tab:ga_mcmc_cost}
\end{table}

\section{Convergence Testing}\label{sec:app:convergence}

\begin{figure}[ht!]
    \centering
    \includegraphics[width=\linewidth]{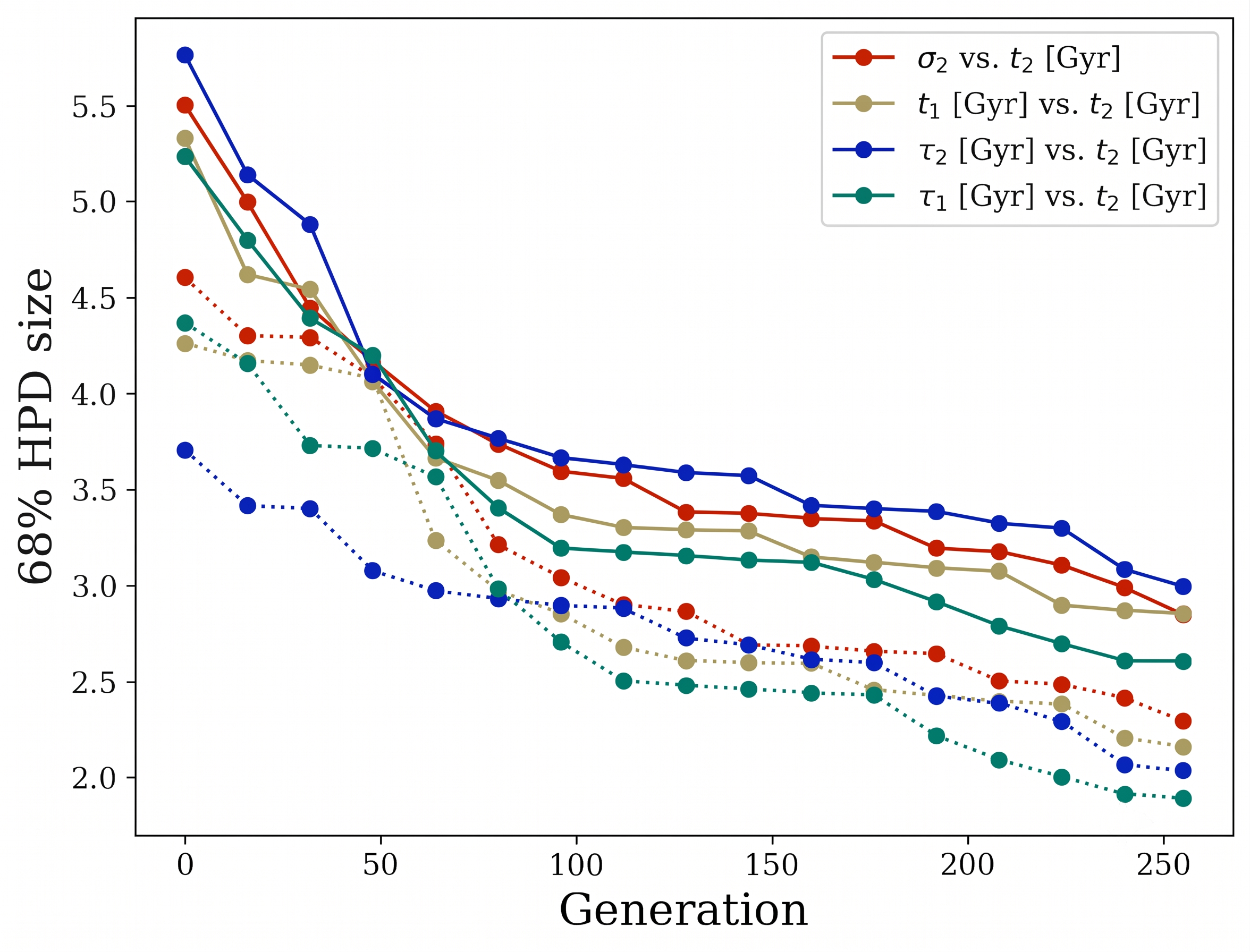}
    \caption{Convergence of the GA posterior. 
    The solid lines show the evolution of the 68\% HPD ellipse size for several parameter pairs as a function of generation for a single GA run.
    The dotted lines show the corresponding HPD sizes measured from the combined pseudo-posterior used in the main corner plot.
    The rapid early decrease and subsequent plateau, together with the close agreement between the solid and dotted curves, indicate the stable and repeatable convergence of the GA.}
    \label{fig:hpd_convergence}
\end{figure}

We assess the GA convergence by tracking the 68\% HPD size of key two-parameter projections as a function of generation.
For each generation, we compute the semimajor axis of the weighted covariance ellipse in the $(\sigma_2,t_2)$, $(t_1,t_2)$, $(\tau_1,t_2)$, $(\tau_2,t_2)$, and $(\tau_1,\tau_2)$ planes using the top-weighted models from a single GA run (solid curves).
As shown in Figure~\ref{fig:hpd_convergence}, the 68\% HPD ellipse sizes shrink rapidly in early generations and then plateau, indicating the stable convergence of the GA.
A second set of curves (dotted lines) shows the same HPD sizes measured from the combined pseudo-posterior used to build the main corner plot in the text (i.e., all GA+DEMC runs together). The dotted curves closely track the solid curves at late generations.
The agreement between the single-run HPD sizes and the combined posterior, together with the monotonic shrinkage and saturation of the HPD envelopes, demonstrate that independent GA realizations converge reliably and consistently to the same region of parameter space.

\clearpage
\bibliography{paper}
\bibliographystyle{aasjournal}

\end{document}